\documentclass[preprint,3p,times,authoryear]{elsarticle}
\usepackage{graphicx}
\usepackage{lineno}
\usepackage{amsfonts,amsmath,amssymb}
\usepackage{color}
\usepackage{natbib}

\journal{Advances in Space Science Research}

\begin{document}

\begin{frontmatter}

\title{THEO Concept Mission: Testing the Habitability of Enceladus's Ocean}
\author[UI]{Shannon M. MacKenzie\corref{cor1}}
\cortext[cor1]{Corresponding Author \ead{s.mackenzie.france@gmail.com}}
\author[BU]{Tess E. Caswell}
\author[UO]{Charity M. Phillips-Lander}
\author[JPL]{E. Natasha Stavros}
\author[Cornell]{Jason D. Hofgartner }
\author[BU]{Vivian Z. Sun}
\author[WU]{Kathryn E. Powell}
\author[UM]{Casey J. Steuer}
\author[CalTech]{Joesph G. O'Rourke}
\author[UCSD]{Jasmeet K. Dhaliwal}
\author[UA]{Cecilia W. S. Leung}
\author[UMD]{Elaine M. Petro}
\author[NAU]{J. Judson Wynne}
\author[SU]{Samson Phan}
\author[UCB]{Matteo Crismani}
\author[MIT]{Akshata Krishnamurthy}
\author[NJ]{Kristen K. John}
\author[JPL]{Kevin DeBruin}
\author[JPL]{Charles J. Budney}
\author[JPL]{Karl L. Mitchell}

\address[UI]{Department of Physics, University of Idaho, Moscow, ID }
\address[BU]{Department of Earth, Environmental, and Planetary Sciences, Brown University, Providence, RI } 
\address[UO]{School of Geology and Geophysics, University of Oklahoma, Norman, OK}
\address[JPL]{Jet Propulsion Laboratory, California Institute of Technology, Pasadena, CA }
\address[Cornell]{Department of Astronomy, Center for Radiophysics and Space Research, Cornell University, Ithaca, NY } 
\address[WU]{Department of Earth and Planetary Sciences, Washington University in St. Louis, St. Louis, MO } 
\address[UM]{Department of Climate and Space Sciences Engineering, University of Michigan, Ann Arbor MI } 
\address[CalTech]{Division of Geological and Planetary Sciences,California Institute of Technology, Pasadena, CA}
\address[UCSD]{Scripps Institution of Oceanography, University of California San Diego, La Jolla, CA }
\address[UA]{Department of Planetary Sciences, Lunar and Planetary Laboratory, University of Arizona, Tucson, AZ} 
\address[UMD]{Department of Aerospace Engineering, University of Maryland, College Park, MD} 
\address[NAU]{Department of Biological Sciences, Northern Arizona University, Box 5640, Flagstaff, AZ 86011 and The SETI Institute, Carl Sagan Center, 189 Bernardo Ave., Suite 100, Mountain View, CA  94043}
\address[SU]{Department of Engineering, Stanford University, Stanford, CA } 
\address[UCB]{Laboratory for Atmospheric and Space Physics, University of Colorado, Boulder, CO }
\address[MIT]{Department of Aeronautics and Astronautics, Massachusetts Institute of Technology, Cambridge, MA } 
\address[NJ]{NASA Johnson Space Center, 2101 NASA Parkway, Houston, TX} 

\begin{abstract}
Saturn's moon Enceladus offers a unique opportunity in the search for life and habitable environments beyond Earth, a key theme of the National Research Council's 2013-2022 Decadal Survey. A plume of water vapor and ice spews from Enceladus's south polar region. \emph{Cassini} data suggest that this plume, sourced by a liquid reservoir beneath the moon's icy crust, contain organics, salts, and water-rock interaction derivatives. Thus, the ingredients for life as we know it-- liquid water, chemistry, and energy sources-- are available in Enceladus's subsurface ocean. We have only to sample the plumes to investigate this hidden ocean environment. We present a New Frontiers class, solar-powered Enceladus orbiter that would take advantage of this opportunity, Testing the Habitability of Enceladus's Ocean (THEO). Developed by the 2015 Jet Propulsion Laboratory Planetary Science Summer School student participants under the guidance of TeamX, this mission concept includes remote sensing and \emph{in situ} analyses with a mass spectrometer, a sub-mm radiometer-spectrometer, a camera, and two magnetometers. These instruments were selected to address four key questions for ascertaining the habitability of Enceladus's ocean within the context of the moon's geological activity: (1) How are the plumes and ocean connected? (2) Are the abiotic conditions of the ocean suitable for habitability? (3) How stable is the ocean environment? (4) Is there evidence of biological processes?  By taking advantage of the opportunity Enceladus's plumes offer, THEO represents a viable, solar-powered option for exploring a potentially habitable ocean world of the outer solar system.
\end{abstract}




\end{frontmatter}

\section{Introduction}

Since the discovery of Enceladus's subsurface ocean \citep[e.g.][]{2006Sci...311.1393P,2006Sci...311.1401S,2008Natur.451..685S,2011Natur.474..620P,2014Sci...344...78I,2016Icar..264...37T}, the small moon of Saturn has been considered a potentially habitable world in the solar system \citep[e.g.][]{2008AsBio...8..909M,parkinson2008habitability,2014AsBio..14..352M}. Subsequent observations and investigations by \emph{Cassini} have revealed that Enceladus's ocean depths are not only liquid, but also warm \citep[e.g][]{2007Icar..187..569M,2015NatCo...6E8604S,2015Natur.519..207H}, salty \citep[e.g.][]{2009Natur.459.1098P,2015NatCo...6E8604S}, and host to a range of interesting organic compounds \citep[e.g.][]{2006Sci...311.1419W,2009Natur.460..487W}. As such, the subsurface ocean is already known to have the three elements identified as necessary for life-- liquid water, chemistry, and energy. Enceladus is therefore of crucial importance to the search for life in our solar system, the study of how habitable environments develop, and defining what ``habitable" might mean. What makes Enceladus most unique, however, are the vents that connect the ocean to the surface, releasing water ice and vapor in geyser-like plumes \citep[e.g.][]{2006Sci...311.1406D,2006Sci...311.1401S,2006Sci...311.1393P,2006Sci...311.1422H,2006Sci...311.1419W,2006Sci...311.1416S}. This readily available ocean material makes testing whether Enceladus's hidden ocean world is habitable both possible and relatively easy with today's technology on a medium-sized mission budget. \\

While water ice is an abundant resource in the outer solar system, liquid water is more unusual. Enceladus's water plumes and subsurface ocean were two of the great discoveries of the \emph{Cassini} mission; a true team effort with lines of evidence from the magnetometer suite \citep{2006Sci...311.1406D}, the Composite InfraRed Spectrometer (CIRS) \citep{2006Sci...311.1401S}, the Imaging Science Subsystem camera (ISS) \citep{2006Sci...311.1393P}, the Ultraviolet and Visible Imaging Spectrometer (UVIS) \citep{2006Sci...311.1422H}, the Ion Neutral Mass Spectrometer (INMS) \citep{2006Sci...311.1419W}, the Cosmic Dust Analyzer (CDA) \citep{2006Sci...311.1416S}, the Visual and Infrared Mapping Spectrometer (VIMS) \citep{2006Sci...311.1425B}, and the Radio Science Subsystem \citep{2014Sci...344...78I}. The plumes emanating from Enceladus's south polar region were directly sampled by INMS and CDA and observed in occultation by UVIS. The results indicated that the plumes are mostly water vapor and ice. The source of that water was identified to be a subsurface ocean using the derived gravity field \citep{2014Sci...344...78I}, libration observations \citep{2016Icar..264...37T}, stress field analysis \citet{2011GeoRL..3818201P}, and the detection of potassium salts \citep{2011Natur.474..620P} in plume-derived \citep[e.g.][]{1981Icar...47...84B,2006Sci...311.1416S,2007MNRAS.377.1588H,2007GeoRL..34.9104J,2008Icar..193..420K,2010Icar..206..446K} E-ring material. The crust of Enceladus is thought to be pure water ice \citep{2006Sci...311.1425B,2007GeoRL..3423203Z} and therefore cannot be the sole source of a salty plume. The plumes thus provide a unique opportunity to understand the internal processes of an icy body with a subsurface ocean. Be they hydrothermal, geochemical, or perhaps even biological, such processes are relatively easy to explore at Enceladus from orbit without the need to land, drill, or rove.\\

Sampling the plumes has also indicated that a rich chemistry is in the subsurface ocean. Trace amounts of organics-- such as methane, ethane, butane, and pentane-- are present in the plume vapor, according to the data from INMS. (See Table 1 of \citet{2009Natur.460..487W} for a comprehensive list.) During higher-velocity flybys, the mass spectrometer observed higher percentages of carbon compounds. This is consistent with longer chain organics breaking into smaller compounds upon impact in the instrument, thus fortuitously facilitating their detection within the limited mass range of INMS \citep{2015Icar..257..139P}. Biologically available nitrogen (nitrogen bearing compounds other than N$_2$) and salts (NaCl, NaHCO$_{3}$, Na$_{2}$CO$_{3}$, and potassium salts) have also been detected \citep[e.g.][]{2006Sci...311.1419W,2009Natur.460..487W,2011GeoRL..3811202H,2011Natur.474..620P}. \\

There are several avenues from which energy can be produced and made available to Enceladus's ocean, including hydrothermal, chemical, and geothermal processes.  Silica spherules were identified by \citet{2015Natur.519..207H} in the E-ring. This discovery strongly supports a warm oceanic environment driven by hydrothermal reactions between the liquid and rocky core. The inventory of organic molecules and other oxidizable species identified by INMS \citep{2006Sci...311.1419W,2009Natur.460..487W} indicates that redox reactions are also a potential energy source. CIRS has observed that most of the 15 GW of thermal emission from the south polar terrain is localized on the ``tiger stripes" \citep{2011JGRE..116.3003H}, the $\sim$130 km long fissures \citep{2015Icar..258...67N} from which the plumes erupt, though exactly what processes drive this geothermal activity is not yet clear. \\

\emph{Cassini} has revealed Enceladus to be a world of true astrobiological potential with evidence for liquid water, chemistry, and energy. However, these results have only scratched the surface. We do not yet fully understand how the ocean is connected to the surface or what drives that activity. Hence, the extent to which sampling the plumes is equivalent to sampling the ocean is unclear. Nor do we know if and how Enceladus could sustain enough heat production to maintain an ocean. It is thus difficult to quantify the lifetime of this liquid reservoir, dramatically affecting which Earth extremophiles would be considered good Enceladus analogs. We do not have a complete inventory of higher order organics, heavy compounds, isotopes, or noble gases present in the plumes. Without this, it is difficult to thoroughly characterize what kind of environment the ocean represents and impossible to quantify whether life is presently at Enceladus. Therefore, we have yet to capitalize on the full scientific potential of sampling Enceladus's plumes.\\

The most exciting question surrounding Enceladus is whether extant life exists. We propose, however, that answering that question alone wouldn't give the complete picture of Enceladus's astrobiological potential. If life is there, what conditions support that biology? How long have conditions been suitable for this life to evolve? What Earth organisms are found in similar habitats? Or, if life isn't there, why not? Could it have been supported in the past? Did the environment change? Fully characterizing this ocean world to answer these fundamental questions about life and where it can develop or survive require a dedicated mission with the latest technology. \\

We therefore present here a concept study for a mission to Enceladus that would follow up on the discoveries of \emph{Cassini}. This paper summarizes how a solar-powered Enceladus orbiter called THEO, Testing the Habitability of Enceladus's Ocean, could specifically determine the state of biologically favorable conditions and search for evidence of biological activity at the small moon. THEO is the result of the 2015 Planetary Science Summer School hosted by the Jet Propulsion Laboratory (JPL) at the California Institute of Technology, the purpose of which is to offer the participants an authentic but primarily educational experience of the mission proposal process. The concept mission was vetted during an intensive week of collaboration with JPL TeamX, resulting in a viable mission adhering to the 2009 New Frontiers Announcement of Opportunity \citep{NFAO09}. We report here our findings of this exercise, that important science (as identified by the planetary community \citep{NAP13117}) is possible with a solar powered mission at Saturn, to contribute to the discussion of future missions. \\

In Section \ref{goals}, we discuss the science mission of THEO, the hypotheses THEO would test and the data THEO would acquire to do so. We then describe the mission architecture (instrument payload, mission design, and spacecraft design) in Section \ref{missionarchitecture} and include a discussion of key trades our team considered. Cost and risk assessment are detailed in Section \ref{management} and we conclude with how Enceladus science can be accomplished with a medium class mission in Section \ref{conclusions}. \\

\section{Mission Goals}
\label{goals}
THEO would address unanswered questions that directly relate to establishing what kind of environment lies beneath Enceladus's icy crust.  There are, of course, countless scientific objectives that \emph{could} be investigated at Enceladus, but success as a medium-sized mission requires limiting our concept mission's scope. Therefore, the overarching theme of THEO seeks to characterize the Enceladus ocean and plumes to determine if lifeforms like those found on Earth could survive or evolve in such an environment. To do this, we must answer the following questions: (1) How are the plumes connected to the subsurface ocean? (2) Are the abiotic conditions habitable? (3) How stable is the ocean environment? and (4) Is there evidence of biological processes?\\

\begin{figure*}
\begin{center}
\includegraphics[width=1\textwidth]{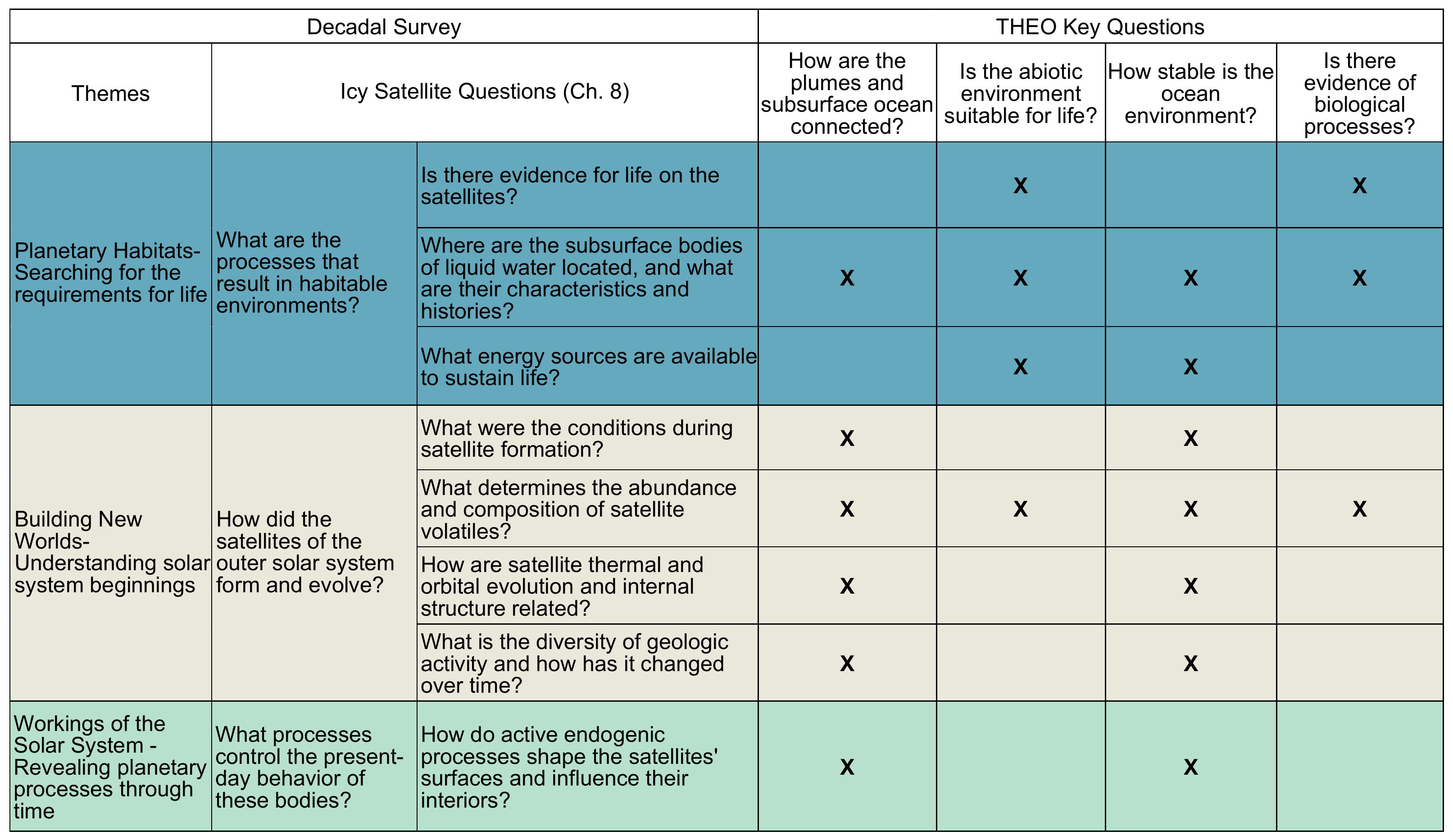}
\caption{Relationship between THEO data and the 2013 Decadal Survey. THEO's main goal, of establishing the habitability of Enceladus's ocean, directly addresses the planetary habitats theme. However, the data from THEO would also address questions from at least two other themes of the 2013 Decadal Survey. \label{DS} }
\end{center}
\end{figure*}

Figure \ref{DS} illustrates the direct connections between these four questions and the themes and areas of interest highlighted in the 2013 Decadal Survey. Our concept mission focuses on one goal: assessing Enceladus's habitability. However, as demonstrated by Figure \ref{DS}, this one goal addresses many of the outstanding questions identified as important by the planetary science community. \\

\begin{figure*}
\begin{center}
\includegraphics[width=0.9\textwidth]{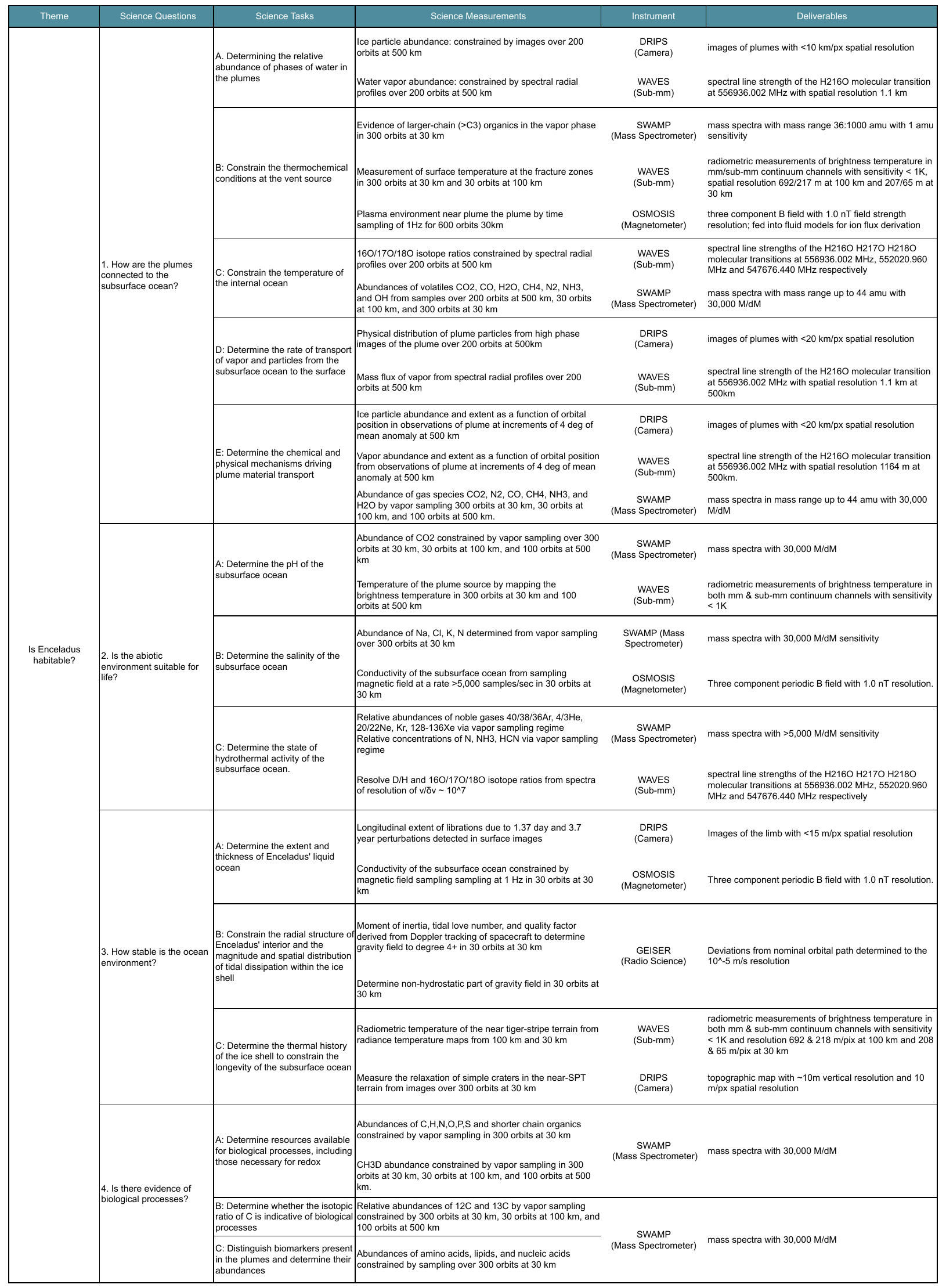}
\caption{The science traceability matrix for the proposed mission. This table lists the science mission theme and questions, the supporting experiments, and what data would be produced to answer the questions. \label{STM}}
\end{center}
\end{figure*}

The proposed mission design approaches the key questions with five instruments as the spacecraft would be equipped with a mass spectrometer, sub-mm instrument, camera, magnetometers, and a radio science instrument. Figure \ref{STM}, the concept mission's science traceability matrix, describes the experiments that would be performed, expected observations from each instrument, and how the results address the mission's four fundamental questions. \\

The linchpin of THEO's mission strategy is providing context for all \emph{in situ} sampling regimes. Proper interpretation of plume samples, including their implications for biological activity and the abiotic conditions of the reservoir, requires knowledge of how material from Enceladus's ocean is driven through the crust to the surface and whether any unique processes take place within the vents. To place these results in the proper timescale, we would investigate the stability of the ocean itself: why it is liquid now, and how it might remain so. By completing these objectives, THEO would form a more complete picture of whether Enceladus could support life.\\

This section further details how THEO would address each of the four key questions listed in Figure \ref{STM}. We describe the proposed science experiments and the data produced, outlining how the results would allow us to discern between hypotheses and models relating to Enceladus's habitability.\\

\subsection{How are the plumes connected to the subsurface ocean?}
\label{Q1}
Unlike other icy satellites with subsurface oceans whose thick crusts prohibit direct sampling with today's technology, Enceladus's plumes offer the unique opportunity for \emph{in situ} exploration from orbit. In order to take full advantage of the plumes, however, it is necessary to identify the mechanisms that transport ice grains and vapor from the ocean to the surface. Understanding this plume-ocean connection would allow us to disambiguate any compounds that are a byproduct of or altered during transport when interpreting plume composition. \\

Several hypotheses have been put forth to explain how the water is delivered from the subsurface ocean to the surface. For example, the thermodynamic model of \citet{2015GeoRL..42.1334B} predicts that clathrates at the ice/ocean boundary of Enceladus's subsurface could participate in plume mechanics. Other models employ an equilibrium boundary between the liquid reservoir and ice grains and vapor at the bottom of the fractures \citep{2008Natur.451..685S,2014AJ....148...45P}. Because the vent structure, wall temperature, and gas densities are largely unknown, the predictions of the plume mechanics models for observable quantities-- such as the abundances and velocities of ejected ice and vapor-- vary. Thus, the THEO mission would address the connection between plume and ocean by measuring the ice-vapor ratio, the chemical composition of the vapor, the spatial distribution and velocities of plume ice grains, and the structure and temperature of the plume vents. \\

The ice-vapor ratio of the plumes is a function of the eruptive process. Clathrate decomposition predicts smaller ratios while equilibrium at a liquid boundary would result in higher ratios. Results from \emph{Cassini} for the ice-vapor content in the plume are inconclusive; the ratio has been reported as 0.42 \citep{2006Sci...311.1393P}, 0.21 \citep{2009Icar..203..238K}, and 0.35-0.71 \citep{2011Icar..216..492I}. (See \citet{2011Icar..216..492I} for a comprehensive discussion of the different analyses.) The camera and sub-mm instrument onboard the THEO spacecraft would capture concurrent images and spectra of the plumes. These combined data sets would yield a more precise measurement of the ice-vapor abundance than has been possible with \emph{Cassini}. Furthermore, the mission design includes repeat coupled observations, ensuring that any spatial and/or temporal differences in the abundance could be observed. Thus, by measuring the ice-vapor ratio and comparing spatial and temporal column abundances, THEO would help identify the mechanism driving Enceladus's eruptions.\\

Vent models rely on the observed plume characteristics (e.g. velocity and size distributions) to investigate rates and mechanics of transport from ocean to surface. For example, vents with smaller volume relative to the surface area in contact with the ocean would erupt gas with higher velocities \citep{2015Natur.521...57S}.  The velocities of the vapor have been estimated by fitting UVIS data \citep{2006Sci...311.1422H,2007Icar..188..154T} to be $\sim$300-500 m/s. \citet{2008Natur.451..685S} used these values (and other parameters) to model vapor condensation and particle growth within the vents. THEO would provide better constraints to such investigations by conducting direct, repeat measurements of the vapor velocity with the sub-mm instrument, which is sensitive to the Doppler shift of the H$_2$O spectra. \\

The spatial distribution of the plume material also reflects the plume-ocean connection. Discrete jets, for example, are inconsistent with a brittle ice shell, while curtain-like eruptions would be indicative of a deep reservoir with a surface area of liquid to chamber ratio large enough to allow the particles and gas enough time to interact \citep{2015Natur.521...57S}. Jets of high velocity ejecta have been documented in ISS data by \citet{2006Sci...311.1393P,2006Sci...311.1422H,2007Natur.449..695S}; and \citet{2008Natur.456..477H}. UVIS observations have also been interpreted as observing discrete areas of higher water density when compared to the broader plume \citep{2011GeoRL..3811202H}. Recently, however, a jets-only interpretation was questioned by \citet{2015Natur.521...57S} who reproduced ISS images of the backlit plumes by modeling plume material as curtains of particles and vapor. These authors argued that the fine structure of a plume curtain can appear as discrete sources in superposition. By taking high phase images of the plumes taken over hundreds of orbits, THEO would be able to map the physical structure of the plumes in higher spatial and temporal detail, allowing the research community to discern between jet- or curtain-like eruptions. Two hundred observations, taken at $\sim$100 m resolution at about the same time of year as the data analyzed by \citet{2015Natur.521...57S}, would represent a seven-fold increase in data over that produced by Cassini thus far. Additionally, as \citet{2013Natur.500..182H} observed a correlation between plume brightness and predicted tidal stresses, these images would be spread over various points along Enceladus's eccentric orbit. Different models \citep[e.g.][]{2015NatGe...8..601B} for the size and structure of plume vents predict different responses to time-varying tidal stresses.  \\

The composition of the plumes could also be constrained by THEO's magnetometer. The deceleration and deflection of the impinging magnetospheric plasma near the plume of Enceladus leaves a measurable imprint on the magnetic field configuration \citep{2006Sci...311.1406D}. Numerical simulations have already been generated to study this behavior \citep{2008GeoRL..3520105S}. Indications to the quantity and behavior of both ions and neutrals near the plume have been made by comparing these simulations with \emph{Cassini} magnetometer data \citep{2011JGRA..11610223K,2011JGRA..116.4221S}. Data from THEO's magnetometers could be used to better characterize these complex interactions with the magnetic field. The B field components (as a function of time and space) are an important input into models like that of \citet{2011JGRA..11610223K} that predict quantities and behavior of ions and neutrals near the plume. These models, in turn, would better our understanding of plume composition and dynamics, necessary for putting other sampling results in the context of either unique to plumes or reflective of the subsurface ocean. \\

While characterizing the content and structure of the plumes themselves, THEO would also investigate the locations from which the plumes erupt. The temperature of the fissures where the plumes originate is a critical parameter in models describing the system. For example, \citet{2010Icar..206..594I} and \citet{2013Icar..226.1128G} predict mass loss rates and calculate radiated power based on observed surface temperatures, the most precise of which comes from a single, high-resolution measurement by Cassini VIMS. These data estimate a vent temperature of 197 $\pm$ 20 K. To discern the thermal budget at the plume source, THEO's sub-mm would map the brightness temperature of the south polar terrain with thermal resolution $<$ 1K at a best spatial resolution of 217 m. At this resolution, THEO would not resolve the width of individual vents, which are likely on the order of ten meters \citep{2013Icar..226.1128G}, but would compliment the CIRS data set \citep{2011JGRE..116.3003H} with similar, if not better, resolution and temperature sensitivity, as well as providing temporal context not possible with \emph{Cassini} data. Measurements taken by THEO would greatly improve the results of thermodynamic modeling by more accurately determining the temperature, monitoring any change over time, and quantifying differences and similarities between fissure temperatures.\\

The temperature of the ocean is also a critical parameter for thermal modeling the ocean-to-plume vents. An inventory of volatiles like CO$_2$, CO, H$_2$O, CH$_4$, N$_2$, NH$_3$, and OH can illuminate conditions at this boundary. The relative abundances of these species indicate the chemical equilibrium at the ocean which depends on temperature. For example, the detection of SiO$_{2}$ by \citet{2015Natur.519..207H} suggests that geothermal activity at the ocean-core boundary drives hydrothermal reactions and that thermally buoyant liquid may interact there. The abundances of numerous other species in the subsurface ocean, including hydrocarbons, ammonia, methane, carbon dioxide, carbon monoxide, and nitrogen, are also determined by temperature \citep{2007Icar..187..569M}.  Oxygen isotopes can also indicate thermal conditions of the ocean as high temperatures favor the formation of certain oxygen-bearing minerals, leading to relatively depleted levels of $^{18}$O in the water vapor produced at the plumes. THEO would be able to address each of these factors using the mass spectrometer and sub-mm instruments. The mass spectrometer would determine abundances for the above mentioned species at three altitudes-- heavier species are thought to fall out of the plume faster, necessitating measurements at high-and low-altitudes to generate a more  complete picture of plume composition \citep{2015Icar..257..139P}. The sub-mm instrument would take spectra at sufficient resolution to distinguish between the oxygen isotopes $^{16}$O, $^{17}$O, and $^{18}$O. \\

With the data provided by the experiments outlined here-- namely, quantifying the abundance and distribution of the plume constituents and the thermal budget of the vents-- thermodynamic models would be able to more accurately explain the behavior of the plumes and predict what alteration might occur during the journey from ocean to surface. Defining precisely how sampling the plumes is related to sampling the ocean would be a crucial insight from the THEO mission. Without it, it would be difficult to determine the habitability of the ocean from the remainder of our experiments that rely on interpreting samples gathered from the plume. \\

\subsection{Is the abiotic environment suitable for life?}
\label{Q3}
Here on Earth, organisms have been found to inhabit a broad range of environmental conditions-- e.g.  pH, salinity, and temperature. Quantifying these abiotic characteristics of the ocean environment is therefore necessary for understanding what kind of life might evolve or survive in Enceladus's subsurface ocean or why terrestrial-like lifeforms cannot. These characteristics also reveal the possible thermal, chemical, and geological processes at work under Enceladus's icy crust. \\ 

pH indicates the acidity or alkalinity of the aqueous environment. While most biological processes take place at circum-neutral pH (5-8), there are unicellular organisms that thrive in highly acidic environments (pH$<$1) and others that prefer very basic environments (pH$>$9) \citep{rothschild2001life}. Current estimates of the pH of Enceladus's subsurface ocean are between 5.7 and 13.5 \citep{2007GeoRL..3423203Z,2009Natur.459.1098P, 2012Icar..220..932M,2015GeCoA.162..202G,2015Natur.519..207H} and were derived from modeling the thermodynamics of species observed in the plumes and E-ring particles by \emph{Cassini}. This range is too large to constrain the ionic potential of the ocean or the sort of life that might inhabit it.\\ 

Two important inputs for thermodynamic modeling of the ocean's pH are the amount of dissolved CO$_2$ and the temperature. The amount of CO$_2$ dissolved in the ocean affects its acidity; increasing the effective concentration of CO$_2$ available for reactions decreases pH. The concentration of CO$_2$ in the plumes-- and, hence, in the ocean-- is thus a critical indicator of chemical potentials in the ocean. \citet{2015GeoRL..42.1334B} and \citet{2015GeCoA.162..202G} use only data from low-velocity flybys of Cassini (a few km/s) to fit the mixing ratio of CO$_2$ to H$_2$O at 0.006 $\pm$ 0.00016 as these measurements are thought to give more accurate gas compositions \citep{2009Natur.460..487W,2015GeCoA.162..202G}. With even lower orbital velocities (between 126-206 m/s), THEO would thus create mass spectra of higher fidelity than any previously obtained. pH is also a temperature-dependent variable, so pH models depend on the ocean temperature and the surface temperature of the vents. Combined with maps of surface temperature from our sub-mm radiometric data, THEO would pave the way for more accurate calculations of the pH of Enceladus's ocean, and therefore a better understanding of what kinds of life might inhabit it. \\

The ocean's salt inventory also provides important insights into biological and geochemical activity. Salts often provide the ions necessary for transportation across cell membranes, but high salinity solutions may negatively impact microbial metabolic processes \citep{stevenson2015multiplication}. The salinity of Enceladus's subsurface ocean will therefore determine the range of habitable conditions and the types of microorganisms that could evolve there \citep[e.g.][]{norton1988survival,rothschild1994metabolic,kamekura1998diversity,seckbach2013enigmatic}. Salt content also indicates which rock-water interactions occur.  Of particular interest is hydrothermal activity as it affects the temperature of the ocean and can constrain the amount of dissolution and precipitation taking place at the water-rock boundary.  Cassini's CDA has observed salt-rich ice particles in the E-ring \citep{2009Natur.459.1098P} and in the plumes \citep{2011Natur.474..620P}. Ground-based observations constrain the salt content in the plume vapor to be at most 0.5-1.0 $\times$ 10$^8$ atoms/cm$^2$ \citep{schneider2009no}. THEO's mass spectrometer would establish the amount of sodium in the vapor with multi-altitude sampling and temporal context to provide more accurate inputs for the Enceladus interior models.  \\

Indirect evidence of the ocean's salinity is also obtainable with THEO's magnetometers. The saltiness of the ocean relates directly to its conductivity, a key parameter of the induced electric and magnetic fields that form as Enceladus orbits Saturn. Because Saturn's magnetic field is aligned with its spin axis, the variations in Saturn's magnetic field are not as dramatic as those at Jupiter. Any induced fields at Enceladus would thus be different from those those observed at Jovian ocean worlds \citep[e.g.][]{kivelson1997europa}. The precision and quantity of data collected by THEO's proposed orbiting magnetometer would be sufficient to constrain the induced magnetic field and predict the conductivity, and thus salinity, of the ocean.   \\

Hydrothermal alteration, with its associated implications for the ocean's temperature, can also be inferred from the abundances of radiogenically-derived isotopes in the plumes. Enrichment in $^{40}$Ar, for example, would indicate that hydrothermal activity has persisted for sufficiently long times to allow the isotope to leach from the rock. Unfortunately, $^{40}$Ar was not well resolved by Cassini \citep{2009Natur.460..487W}. THEO's mass spectrometer would be able to identify the isotopic ratios of noble gases, especially $^{40}$Ar/$^{38}$Ar/$^{36}$Ar, Kr, $^{4}$He/$^{3}$He, $^{22}$Ne/$^{20}$Ne, $^{128}$Xe/$^{136}$Xe. These improved measurements would constrain what hydrothermal activity occurs at the water-rock boundary. \\

Hydrogen and oxygen isotopes can also be used to search for evidence of hydrothermal alteration. The relative abundance of deuterium in methane and oxygen isotopes in carbon dioxide isotopologues is affected by whether the species are produced via serpentinization, the addition of water into the rock structure \citep{mousis2009formation}. The mass spectra compiled by THEO's mass spectrometer would be sensitive enough to distinguish between D/H isotopes in CH$_4$ and oxygen isotopes in CO$_2$, thus shedding light on whether abiotic processes play a role in the formation of molecules of biological interest. The proposed sub-mm data would be able to determine the D/H ratio. Combined with radiogenically-derived isotope measurements, THEO would thus provide two independent means of investigating whether hydrothermal activity occurs at the water-rock interface.\\

Finally, the abundance of molecular hydrogen in the plumes can also be used to infer the state of hydrothermal reactions. Hydrothermal systems like Earth's Lost City \citep{kelley2001off} are rampant with molecular hydrogen as a byproduct of several serpentinization reactions \citep{2005Sci...307.1428K}. It is difficult to discern molecular hydrogen in the mass spectra of INMS as  the source of molecular hydrogen detected is degenerate: it is unclear whether the mass abundance observed is from the material sampled in a given flyby or a remnant of high velocity plume sampling. \citep{2015Icar..257..139P}  The mass spectrometer on board THEO would conduct sampling at high enough resolution and low enough relative velocity to resolve the abundance of H$_2$.  The observed value could be compared to models \citep[e.g.][]{2016LPI....47.2885G} that predict the production rate of H$_2$ as a function of geochemical and geophysical conditions.

THEO's inventory of molecules and isotopes present in the plumes would also shed light on the thermal conditions in the ocean. 
Because NH$_3$ depresses the freezing point of water, its abundance in the plume vapor (as well as that of N$_2$, a decomposition product of NH$_3$) would offer another constraint on the ocean temperature \citep{2007Icar..187..569M,2009Natur.460..487W,2015NatCo...6E8604S}. Another example of how species abundances can provide constraints on the temperature of the ocean is studying the HCN abundance. HCN quickly reacts in warm water to produce formic acid and ammonia, the rate of which is in part a function of temperature \citep{miyakawa2002cold}. Thus, as noted by \citet{2009Natur.460..487W}, comparing the abundances of HCN and N$_2$ can shed light on whether this kind of reaction is happening now or happened in the past. \\

Together, THEO's investigation of abiotic environmental factors like acidity, salinity, and temperature would improve the current characterization of Enceladus's ocean by reducing uncertainties on these parameters that are presently too large to make definitive statements on habitability. While characteristics like pH, salinity, and temperature are not tell-tale signs of life, these data would constrain which analog terrestrial environments could guide the search for bio-signatures at Enceladus \citep{preston2014planetary}. Understanding the present conditions of the ocean is also an essential starting point for investigating whether Enceladus's ocean maintains these conditions and thus over what timescales life might have to evolve.\\

\subsection{How stable is the ocean environment?}
\label{Q4}
Critical to the habitability of Enceladus's ocean is its longevity. The energy source powering geologic activity on Enceladus has been investigated since \emph{Voyager} revealed a young surface on the small moon \citep{1982Sci...215..504S}, but  \emph{Cassini's} discovery of the plumes has exacerbated the problem. The observed heat flux from the south polar region is significantly greater than predicted by models for Enceladus's evolution. Cassini CIRS measurements indicate that the endogenic power of the south polar terrain is 15.8 $\pm$ 3.1 GW \citep{2011JGRE..116.3003H}. Radiogenic heating, however, is estimated to provide only 0.3 GW \citep{2007Icar..188..345S} and tidal heating has not yet been shown to accommodate the difference. Several models have been proposed to explain the discrepancy between observed and predicted heat flux such as episodic activity at Enceladus with alternating periods of net energy accumulation and expulsion \citep[e.g.][]{2012Icar..219..655B,2013AREPS..41..693S}. Before firm predictions can be made, however, more information is required. Understanding the energy sources powering Enceladus's plumes and maintaining its ocean is necessary for determining the ocean stability. This in turn provides context for any biological findings, or lack thereof (perhaps the ocean hasn't been liquid long enough for life to evolve, etc.). To this end, THEO would conduct imaging, thermal mapping, and radio science campaigns over the south polar terrain to explore ice shell properties as they relate to heat production and the current extent of the liquid ocean. \\

A variety of mechanisms have been put forth to explain Enceladus's heat flux. Decoupling the ice shell from the silicate interior, for example, increases the amount of heat generated by tidal dissipation \citep{2008GeoRL..35.9201R,2008Icar..196..642T,2010JGRE..115.9011B}. \citet{2014Sci...344...78I} found a gravitational anomaly at the south pole consistent with a large body of liquid that would decouple the south polar terrain, while recent work by \citet{2016Icar..264...37T} identified evidence of librations (longitudinal oscillations of the surface) in the ISS data set that are consistent with a fully decoupled ice shell.  These results were surprising considering calculations that a global ocean would freeze in 100 Myr without transient heating events \citep[e.g.][]{roberts2008near}. \citet{2016Icar..264...37T} were able to place an upper bound on the amplitude of the libration, but greater imaging resolution is required to fully constrain the libration and therefore the structure of the ice shell and extent of the ocean layer. THEO would address this need through imaging observations. \citet{2010GeoRL..37.4202R} predict minimum libration amplitudes on the order of 100 m in the case of full ice-rock coupling, with significant increases in amplitude if the ice shell is fully decoupled. Such amplitudes would be measurable by comparing observations from THEO's camera. Thus, the mission would be capable of confirming the existence of a global ocean and better constraining its structure and the ocean lifetime.\\

The thermal history of the ice shell can also be explored through its rheology. Consider a warmer, less-viscous shell. It is more likely to undergo convection and cool rapidly, but a less-viscous ice shell dissipates more tidal energy to heat the interior. THEO would distinguish between these different possibilities with a variety of methods. Topographic maps would be derived from high-resolution stereo imaging of Enceladus's south polar terrain. These maps would facilitate models of viscous relaxation of observed features that would provide a proxy for the viscosity of the ice shell and thus its thermal history. For example, \citet{2012GeoRL..3917204B} examined crater relaxation in two regions of Enceladus's northern hemisphere and found that the viscous relaxation indicated an average heat flux of 150 mW/m$^2$.  \citet{2010Icar..208..499B} estimated the age and viscosity of a folding layer in the south polar terrain from the fold wavelength, but the scope of that study was limited by the available image resolution. THEO's images (with a best spatial resolution of 10 m/pixel) would be used to validate these estimates and investigate whether shorter wavelength folds are present. Thus, THEO's proposed image data set would allow exploration of the spatial and, in some cases, temporal evolution of the thermal budget of Enceladus's ice shell, illuminating the conditions contributing to maintaining Enceladus's ocean. \\

Finally, geophysical properties such as moment of inertia, tidal Love number, and quality factor are integral to determining the thermal budget of Enceladus. The tidal Love number and quality factor are important parameters required for modeling tidal dissipation, but are not well constrained for Enceladus, leaving the amount of heat generated by tidal dissipation a crucial unknown. Precise measurement of the moment of inertia, for example, would reveal the size and density of the core. A low density core made of hydrated silicates, for example, could suggest that heat-producing $^{40}$K leached into the overlying ocean \citep[e.g.][]{engel1994silicate}. Additionally, determining the hydration state of the core would provide major constraints on whether rock is currently available for hydrothermal reactions. \citep[e.g.][]{2016LPI....47.2885G}.  \\

Mass distribution within Enceladus can be explored via radio science. Variations from a uniform distribution affect the moon's gravity field, slightly perturbing  the spacecraft's orbital velocity. These Doppler shifts are measurable by tracking the signal from the spacecraft's high gain antenna with the Deep Space Network. As Cassini could only conduct flyby observations, the current estimate for Enceladus's gravity field is determined to degree 3 \citep{2014Sci...344...78I}. Spacecraft interactions with the plume and the lack of spherical symmetry of Enceladus complicate the interpretation of gravity data. But because THEO would be an orbital mission, the frequency and coverage of measurements would complement and improve our understanding of Enceladus's gravity field to at least degree 4. Additionally, frequent measurements of the plume gas velocity by the sub-mm instrument and density by the camera would better constrain the neutral particle drag on the spacecraft, making derivations of the changes in the spacecraft's line-of-sight velocity more accurate. Solving for a higher degree, more precise gravity field would provide more accurate parameters to models like those of \citet{2014Sci...344...78I} and \citet{baland2015obliquity} which produce estimates for the tidal Love number, quality factor, and principal moments of inertia.  \\

By quantifying the ice shell characteristics and thermal flux at unprecedented resolution in space and time, THEO could offer a better understanding of just how long the ocean has been and will remain liquid. This is a critical parameter for putting the results of biological and environmental ocean characterization into context. First, as Enceladus is one of several icy ocean worlds in our solar system, these findings would be of interest with respect to formation timescales of other potentially habitable oceans, such as at Europa or Titan. Second, if THEO were to detect life, constraining the time scale of Enceladus's ocean offers an important comparison with how long it took life to evolve on Earth. Or, if we do not detect any biomarkers, the stability of the ocean might explain such an important, albeit null, result. \\

\subsection{Is there evidence of biological processes?}
\label{Q2}
Evidence from \emph{Cassini} suggests that Enceladus is a good candidate for a habitability study, but these data cannot reveal if Enceladus is currently inhabited. A new mission is required to search for evidence of biological processes. With improved resolution and sensitivity, the THEO mass spectrometer would be able to follow up on the insights gained from INMS to do exactly that. THEO has several experiments to test for biological activity, including identifying the presence of biomarkers in the vapor and the relative abundances of C, H, N, O, P, S, organics, and C isotopes.\\

Some of the simplest biomarkers for terrestrial life are amino acids, lipids, and nucleotides. These polymers are essential for biological processes and their presence within the plumes would be strong evidence for life within Enceladus's ocean. These biomarkers are also extremely large, with masses of over 100 amu ($\sim$100 amu for amino acids, $>$ 200 amu for lipids, and $>$ 400 amu for nucleotides).  Cassini's INMS is only sensitive up to 100 amu and therefore unable to resolve any of these biomarkers. THEO's mass spectrometer, sensitive to masses $>$ 1000 amu, would not only detect the mass signature of amino acids, lipids, and nucleotides, but also distinguish them from other macromolecules, if they are present.\\

For THEO to sense these polymers, however, they must be present in the vapor component of the plumes. As of yet, there is no evidence to exclude the possibility of complex organics in the vapor. Recently, however, the mass signatures in CDA data indicate the presence of these target molecules in the ice grains \citep{postberg2015refractory}.  With relatively slow orbital velocities, THEO would not have enough kinetic energy to break apart ice grains upon impact within the mass spectrometer as INMS did in several flybys \citep{2015Icar..257..139P}. In Section \ref{SWAMP}, we discuss how to expand the sample range. However, if biomarkers are present in the plume vapor at the same abundances identified in meteorite samples (amino acids at 60 ppm; adenine, guanine, and uracil at 1.3 ppm in the Murchison meteorite \citep{cronin1986amino}), the planned plume sampling  (see Figure \ref{obsschedule} and Section \ref{missiondesign}) would be sufficient to characterize their abundances.  \\

Even if these ``smoking gun" biomarkers are not observed, there are less direct lines of evidence that THEO could investigate.  Life on Earth has developed using a small set of chemical building blocks: C, H, N, O, P, and S \citep{mckay2004life}. Thus, in an environment with active biological processes, these ``legos" are sequestered at greater concentrations in specific molecules than if only abiotic processes were responsible \citep{mckay2004life,2014AsBio..14..352M}. Specifically, when abiotic processes are responsible, the distribution of the six elements would be more uniform across possible molecules. Consistently high abundances of bio-relevant molecules with C, H, N, O, P, and S would thus be evidence for Earth-similar life processes. Or, if other element combinations are found at high abundances within specific molecules, this may indicate biological processes analogous, not identical to, terrestrial biology.  \\

Data from THEO would vastly improve the available inventory of material in the plume vapor. The mass spectrometer on-board THEO would catalog species of atomic mass between 0-1000 amu with 30,000 M/dM resolving power, a larger and therefore more complete sampling than presently available of the full range of species present in the plume. Species of larger masses than CO$_{2}$ and H$_2$O are more likely to be observed at the lowest observing altitude of our mission (30 km) at some $<$1\% abundance \citep{2009Natur.460..487W} as mass is a driving factor in the distribution of ice grains in the plume \citep{2015Icar..257..139P}. INMS observed species with as low as one in 10$^{-6}$ abundance \citep{2006Sci...311.1419W} during a flyby with closest approach to the plumes of $\sim$ 400 km. With 300 orbits near this altitude (500 km), THEO would collect data over the extent of the tiger stripes, the densest part of the plume \citep{2006Sci...311.1419W} for over 430 hours. At the highest sampling rate, this would translate to over $10^8$ samples. At lower altitudes not sampled by INMS, the density of the plumes is likely orders of magnitude larger. If plume density roughly falls off as 1/r$^2$, at 30 km, water abundance should be on the order of 10$^{20}$m$^{-3}$ and higher order organics observed by INMS would be at least 10$^{14}$m$^{-3}$, well within the 1 ppb sensitivity of SWAMP.   \\

To put the observed inventory of higher-order organics into proper context, it is necessary that THEO discern whether the observed species are biologically-derived products. We know that not all organics are the result of biology. A plethora of exotic, long-chained organic species are formed in the photolytic processes of Titan's upper atmosphere, while Fischer-Tropsch synthesis and thermal degradation create methane from the decomposition of other organic material on Earth and elsewhere in the solar system \citep[e.g.][]{hindermann1993mechanistic,mccollom1999abiotic,mccollom1999methanogenesis,hill2003catalytic}. Biological methanogenisis has been proposed for the methane observed in Enceladus's plumes \citep{2014AsBio..14..352M}, though other works propose that the methane is thermogenic in origin \citep{2007Icar..187..569M}. While not entirely conclusive (see, for example, the work of \citet{allen2006mars} and \citet{horita1999abiogenic}), isotopes offer a potential means of distinguishing between the two processes. Work by \citet{1999OLEB...29..167M}, \citet{sassen2004free} and \citet{proskurowski2008abiogenic} demonstrates that abiotic processes produce a power-law relationship between the concentrations of C$_1$ relative to C$_2$+C$_3$ and the isotope ratio in CH$_4$.  Biological and abiotic processes produce similar concentrations of C$_1$ isotopes, but orders-of-magnitude difference in the concentrations of higher-order C isotopes in methane. \\

Thus, an inventory of hydrocarbons of sufficient resolution to distinguish between C isotopes would inform our understanding of where Enceladus's methane comes from: abiotic chemical reactions, biotic processes, or a mixture of both.  THEO's mass resolution would be sufficient to distinguish between $^{13}$CH$_2$D, $^{13}$CH$_4$, and CH$_3$D, as well as C and H isotopologues. The sub-mm instrument would also be able to distinguish between H isotopologues, enabling inter-instrument comparison in real time and adding redundancy and reliability to our science measurements. \\

Should any biomarkers be found in the plume vapor, these proposed experiments would offer the first direct evidence of another inhabited world in our solar system. However, even if amino acids, lipids, or nucleotides are not observed, THEO would constrain the likelihood of biological activity from isotope and C, H, N, O, P, and S abundances. These two results alone would address the very heart of one of the 2013 Decadal Survey themes: searching for the requirements for life and life itself.\\

\section{Mission Architecture}
\label{missionarchitecture}
The THEO mission architecture was designed via the concurrent systems engineering process of JPL TeamX that facilitates real-time trades between mass, power, volume, cost, and data constraint rates for mission development. Through this process, the JPL Planetary Science Summer School produces mission concept studies of similar caliber to competed missions. We developed an architecture for THEO that best enables the science goals described in the previous section. This New Frontiers class mission would include a five-instrument suite on a solar-powered bus and would conduct an altitude-dependent observing campaign during nearly 1000 orbits of Enceladus ($\sim$ 6 months).  \\

\subsection{Instruments}
\label{instruments}
THEO's suite of instruments would conduct remote sensing and \emph{in situ} experiments. The proposed mission design uses heritage instruments (hardware that has been proven on previous missions or has been selected to fly on missions in the near future) to reduce cost and development time. All estimates for cost, mass, and power are based on the heritage instruments, summarized in Table \ref{insttab}.\\

\begin{table*}
\begin{tabular}{l p{6cm} p{3cm} c c c p{2cm}}
Acronym & Name & Instrument & Data & Mass & Power & Heritage \\
& & & (Gb) & (kg) & (W) & \\
\hline
SWAMP & Space-borne Water Analysis by Molecule Pulverization &Mass spectrometer& 12& 15.3& 50& MASPEX\\
WAVES & WAter Vapor Emissions Sub-mm & Sub-mm & 6 & 22.9 & 59 & MIRO\\
DRIPS & Dynamic Resolution Imaging of the Plumes and Surface & Camera & 186 &1.53 & 12 & Malin Space Science Systems\\
OSMOSIS & Ocean Sensing Magnetometer Orbital Salinity Induction Science & Magnetometers & 1 &6.12 & 3 &numerous \\
GEISER & Gravity Engaging Investigation Sensing Enceladus with Radio & Doppler tracking & - & - & - & numerous 
\end{tabular}
\caption{Summary of the proposed instrument suite of THEO. Each instrument would contribute to the science objectives via the tasks listed in Figure \ref{STM}, but here we list the mission architecture characteristics of each. \label{insttab}}
\end{table*}

\subsubsection{Mass Spectrometer}
\label{SWAMP}
The mass spectrometer, Space-borne Water Analysis by Molecule Pulverization (SWAMP), would measure element and isotope ratios as well as molecule abundances. SWAMP is modeled after MASPEX, the mass spectrometer selected for the Europa flagship and proposed as a part of ELF \citep{2015LPI....46.1525L}. With an extended mass range ($>$1000 amu), mass resolution ($>$30,000 M/dM), and sensitivity (1 ppt), the capabilities of this instrument are sufficient to complete the science tasks of our mission. For comparison, Cassini's INMS is sensitive to a mass range of up to 100 amu at 100 M/dM resolution. SWAMP thus represents a significant enhancement over the capability of our most recent exploration of Enceladus. \\ 

\emph{In situ} sampling would be accomplished by flying THEO through the plumes with SWAMP pointed in the ram direction. In this orientation, vapor would fly into the open and closed sources of SWAMP before being ionized and directed towards the detector to record the time of flight. Because time-of-flight is a function of mass, the data would then be converted into charge versus mass distributions \citep[e.g.][]{waite2004cassini,2015PSS..117..436H} and used to determine abundances of elements and compounds of interest, including isotopic ratios. \\

In orbit, THEO would be able to take an unprecedented amount of data at three altitudes for a comprehensive sampling of the plume vapor content with a state-of-the-art, high resolution mass spectrometer. The relatively slow orbital velocities would remove the problems known to affect INMS of molecules dissociating within the instrument \citep{2015Icar..257..139P}. However, this also means that the spacecraft does not have enough relative kinetic energy to passively break down, and thus sample, ice grains in the plume. Thus, SWAMP (as currently designed) is limited to sampling only the vapor component of the plumes. The instrument on which we based our mass spectrometer is not specifically equipped to actually ``pulverize" ice grains in the plume, but modifications (such as the addition of a filament) may be added to actively break down ice grains before they enter SWAMP. Such modifications (or investigations into other modes of passive breakdown) were beyond the scope of this study, but we note that this would be an important avenue for research to prepare for a THEO-like mission. Regardless, THEO's observing schedule is more than sufficient to resolve and characterize any salts, amino acids, and larger order hydrocarbons found in the plume vapor. \\

\subsubsection{Sub-mm}
\label{WAVES}
The sub-mm instrument, WAter Vapor Emissions Sub-mm (WAVES), would analyze the water vapor content in the plumes at resolutions high enough to distinguish H and O isotopes.  The instrument would be sensitive to the molecular transitions of H$_2$O isotopologues at 556936.002 MHz, 552020.960 MHz, and 547676.440 MHz.  Additionally, this instrument is sensitive to Doppler shifts in water spectra, allowing for the determination of ice grain velocities in the plumes. WAVES is modeled after MIRO flown on Rosetta, a 30-cm diameter telescope and two receivers operating at frequencies of 190 GHz (1.6 mm, the ``mm" component) and 562 GHz (0.5mm, the ``sub-mm" component) \citep{2007SSRv..128..561G}.  These components have 6.9 $\mu$rad and 2.1 $\mu$rad fields of view, respectively, in the spectra-collecting mode. WAVES data could be used to map the spatial extent and strength of the water content in plumes and thus be used to derive vapor content and velocity. Based on the spectral resolving power of MIRO (2x10$^6$, \citet{2007SSRv..128..561G}), WAVES should be able to resolve differences in ice grain velocities on the order of 200 m/s, on the order of estimates for the bulk vapor velocity \citep[300-500 m/s,][]{2006Sci...311.1422H,2007Icar..188..154T} and twice that for ice grains \citep[80-180 m/s,][]{2009ApJ...693.1749H} derived from VIMS data, though this observed range is not sensitive to larger velocities due to the resolution and field of view. 

In the thermal mapping mode, WAVES would measure the thermal emission of the south polar terrain at 218 m/pixel (orbiting at 100 km) and 65 m/pixel (at 30 km) and cover the entirety of the south polar terrain with $<$ 1K resolution. Black body emission curves would then be fit to the data to derive temperature maps and heat production rates. The thermal maps would be comparable to those made by Cassini CIRS, which covered the entire south pole at 6-10 km/pixel.\\

\subsubsection{Camera}
\label{DRIPS}
The Dynamic Resolution Imaging of Plumes and Surface (DRIPS) is THEO's camera, the proposed roles of which include both science tasks and optical navigation. With a design based on previous cameras produced by Malin Space Science Systems, DRIPS would collect data essential to determining the plume ice-vapor ratio, constraining libration amplitudes, and investigating the thermal history of the ice shell. 
By imaging the plumes at high phase in the visible band at an order of magnitude higher resolution than possible with ISS, THEO would create a data set of radiance images that would be converted to I/F maps. \citet{2006Sci...311.1393P} and \citet{2011Icar..216..492I} have demonstrated that the scattering from the plume ice grains at visible wavelengths with a mean radius of 3.1 $\pm$ 0.5 $\mu$m \citep{2011Icar..216..492I} is sufficient to inverse model the reflectance as a function of particle size-frequency distribution, yielding the column abundance.  Cassini measured the total mass of the plume to be 1.45 $\pm$ 0.5 x10$^5$kg \citep{2011Icar..216..492I}. By increasing the integration time and optimizing the phase angle (which is possible with the large number of planned observations), THEO would be capable of detecting a plume many orders of magnitude less massive than observed by Cassini.
These data, taken at a spacing of about 4$^{\circ}$ true anomaly, also represent an improvement in the frequency of observation to characterize plume particle distribution dependence on Enceladus's eccentric orbit. While observing at 30 km, the camera would map the south polar terrain up to 50$^{\circ}$S with 10 m/px spatial resolution. Cassini ISS created regional maps of 110 m/pixel, with select south polar terrain images at up to 7 m/pixel. Repeat imaging with DRIPS would be used to generate stereo maps of the surface with up to $\sim$10 m vertical resolution (on the order of the vertical resolution of ISS, but with more complete coverage of the south polar terrain). These data sets would be used to identify librations of the surface and investigate surface geology. \\

\subsubsection{Magnetometers}
\label{OSMOSIS}
Ocean Sensing Magnetometer Orbital Salinity Induction Science (OSMOSIS) includes two magnetometers, each located at the end of a solar array wing. These instruments would measure B fields in the Enceladus environment by sampling at a rate of 1 Hz. OSMOSIS would determine the three component periodic magnetic field with 1.0 nT resolution, a significant improvement over Cassini MAG in terms of resolution (40 nT) and spatial and temporal frequency of data collection (30 orbits of data with OSMOSIS; 22 flybys with MAG). Processing these data would disambiguate unique fields of the Enceladus environment, such as those of plume interactions and the conducting ocean. The strengths of these components could then be used to determine the conductivity and depth of Enceladus's ocean as well as serve as important inputs to fluid models that predict the ion flux in the plumes.\\

\subsubsection{Gravity Science}
\label{GEISER}
As used on many missions, THEO's high gain antenna (HGA) would also serve as a science instrument by conducting Doppler tracking of the radio signal during contact with the Deep Space Network and is thus known as Gravity Engaging Investigation Sensing Enceladus with Radio (GEISER). The data taken during these experiments would include line-of-sight velocity changes with precision 10$^{-5}$ m/s from which the gravity field could be derived. Despite having the same absolute precision as Cassini, THEO would be roughly an order-of-magnitude more sensitive to velocity perturbations induced by Enceladus because of its relatively slow orbital velocities. For example, the velocity perturbations associated with the degree-3 zonal harmonic coefficient J$_3$ are estimated as ~6 mm/s at an orbital altitude of 100 km, compared to ~0.2 mm/s during a Cassini flyby \citep{2014Sci...344...78I}. While the HGA's operations role would be continuous through the mission, we only list those orbits when the instrument would be used for its scientific tasks in Figure \ref{obsschedule}. \\

\subsection{Mission Design}
\label{missiondesign}
\begin{figure}
\begin{center}
\includegraphics[width=0.7\columnwidth]{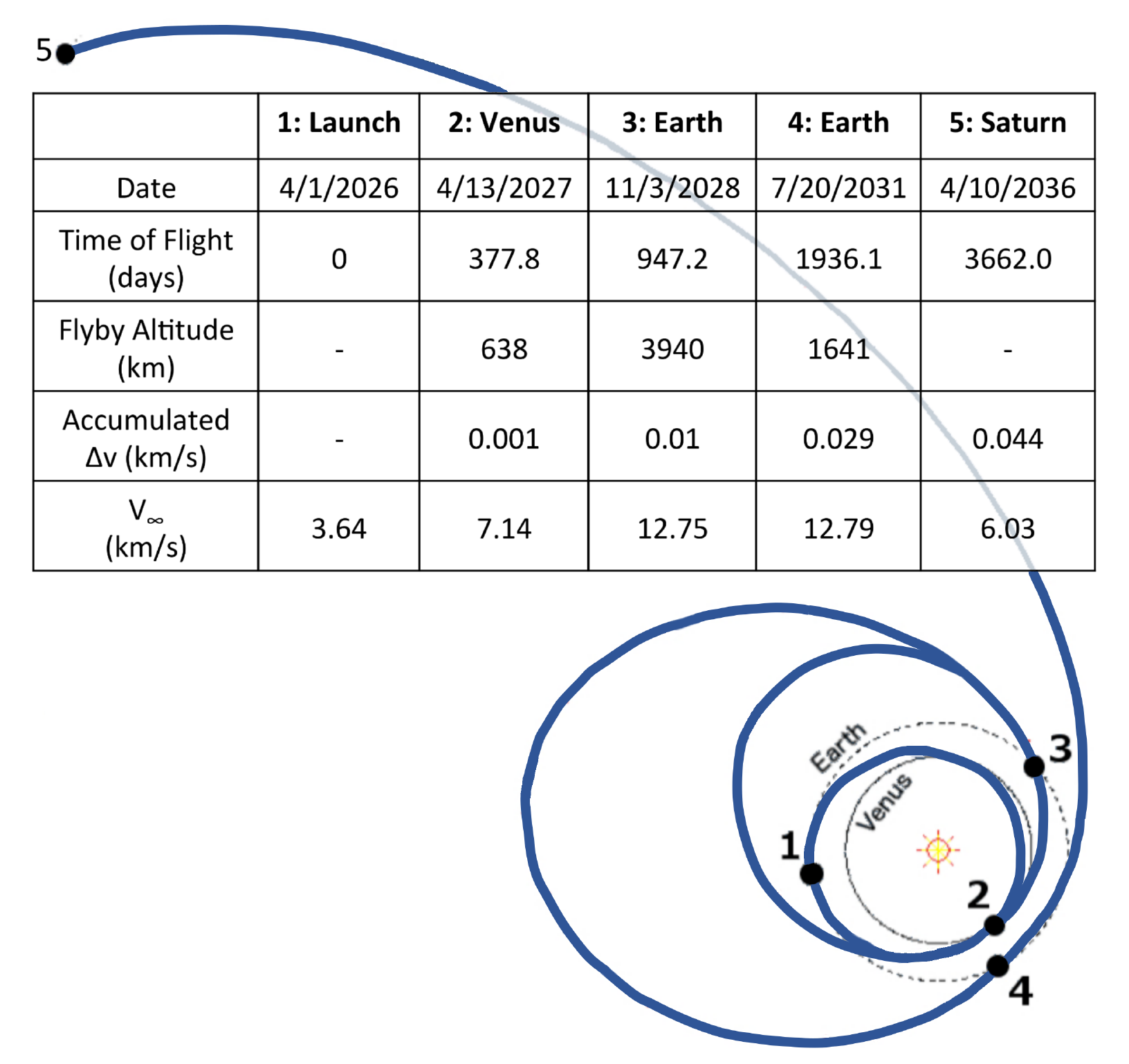}
\caption{Possible inner solar system tour schedule. With no Jupiter gravity assist available, THEO would make use of one Venus and two Earth flybys to get the spacecraft. While in the inner solar system, the spacecraft would operate in standby mode to reduce communications and staffing costs. The potential heating concerns during THEO's proximity to the Sun would be addressed by using the high gain antenna as a heat shield (see Section \ref{risk}). Time of flight is listed in Earth days. V$_{\infty}$ is the hyperbolic velocity of the spacecraft, i.e. as if it were orbiting at an infinite distance from the target. 
\label{eveetour} }
\end{center}
\end{figure}

The THEO spacecraft would launch on an Atlas V 541. As a Jupiter gravity assist was not available during the proposed timeframe of the mission (summarized in Figure \ref{eveetour}), the spacecraft would first complete an inner solar system tour, summarized in Figure \ref{eveetour}, to gain the necessary $\Delta$v to reach Saturn's orbit. The suggested flight path includes one gravity assist from Venus and two from Earth. In total, the journey to the Saturnian system would take ten years, during which no science collection is planned.\\

\begin{figure*}[h!]
\begin{center}
\includegraphics[width=1\textwidth]{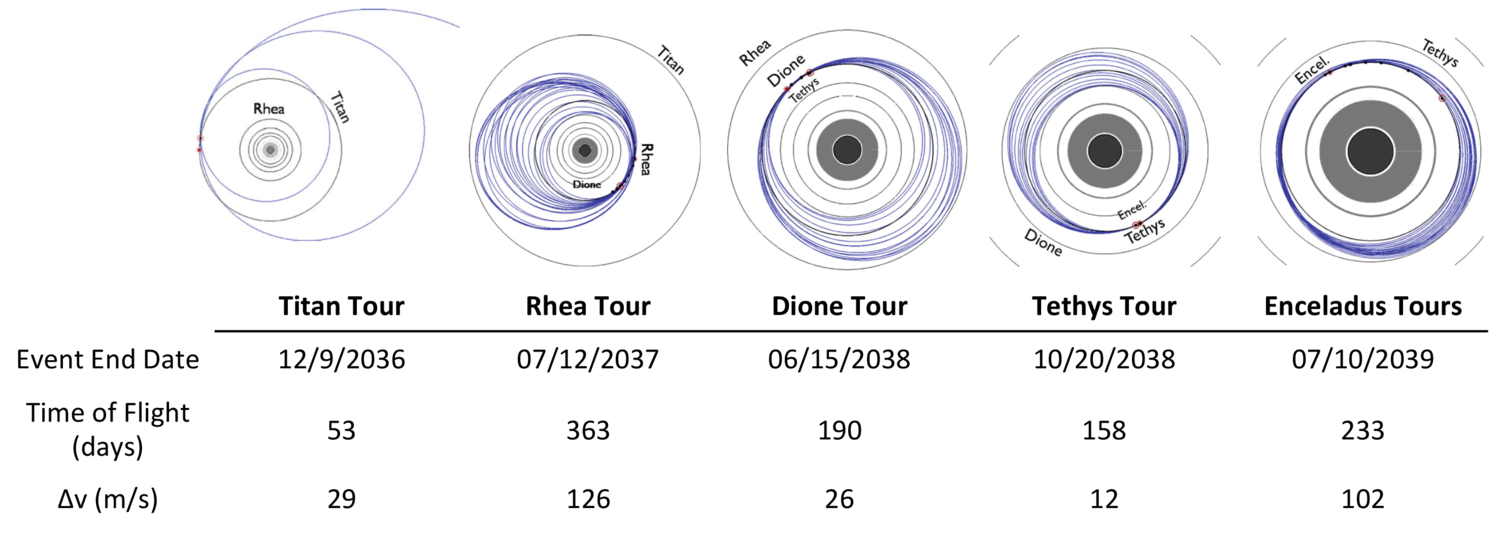}
\caption{Proposed flybys at the Saturn system necessary for Enceladus orbit insertion. No science operations are proposed during this tour of Saturn moons. Though they could increase scientific return, such operations are not within the scope of the proposed mission and would have to be evaluated post-selection. 
\label{saturntour}}
\end{center}
\end{figure*}

The spacecraft would be inserted into Saturn's orbit with a $\Delta$v of 712 m/s. Because of Enceladus's low mass and the location of its orbit deep within Saturn's gravity well, designing a $\Delta$v-efficient rendezvous with Enceladus is a challenge. To reduce the spacecraft's kinetic energy to that appropriate for Enceladus orbit insertion, THEO would conduct a 2.7 year tour of the Saturn system, making a total of 62 flybys of several other Saturnian moons \citep[e.g.][]{strange2009leveraging,2010CeMDA.108..165C}. The schedule is shown in Figure \ref{saturntour} and includes flybys of Titan (3), Rhea (15), Dione (10), Tethys (12), and Enceladus (12). The costed mission plan does not budget for science operations during this phase, but it does not preclude opportunities for mission enhancement post-selection.\\

\subsubsection{Observation Schedule}
\begin{figure*}[H!]
\begin{center}
\includegraphics[width=1\textwidth]{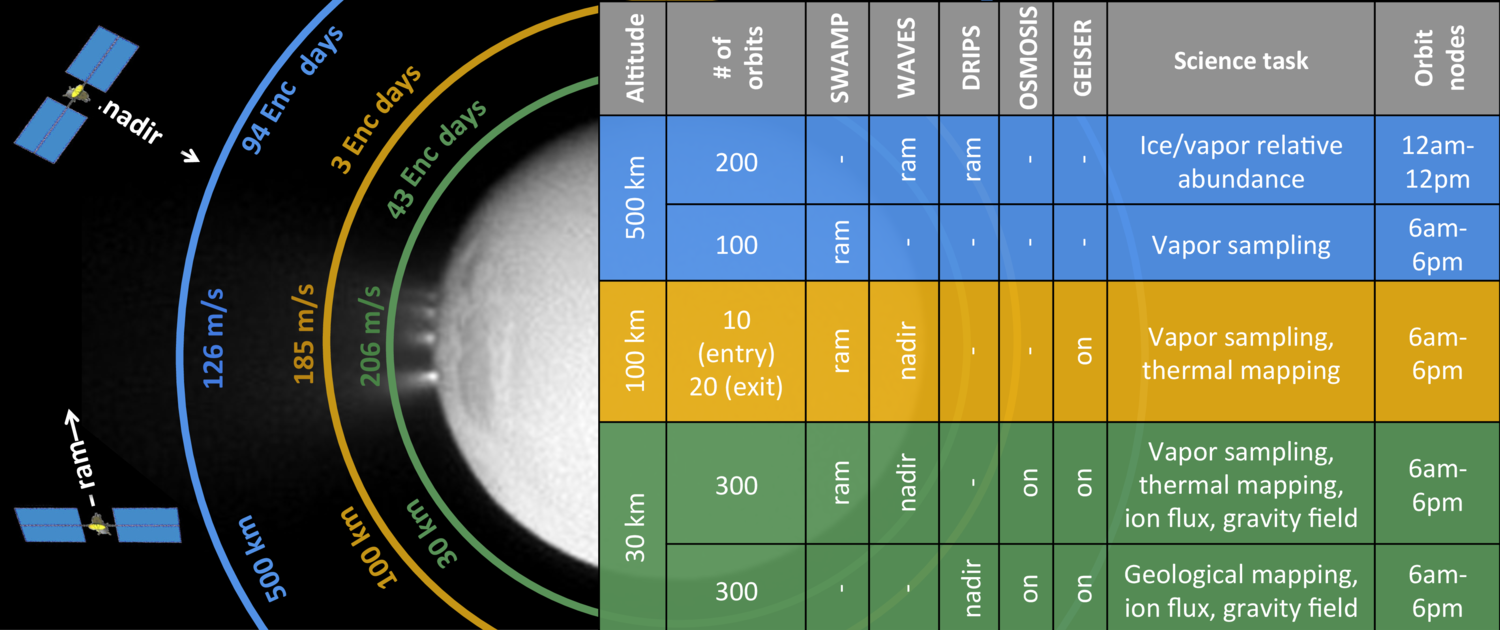}
\caption{Proposed THEO observing schedule. The altitude, spacecraft velocity, and total time spent in orbit at that altitude (in Enceladus days, 1 Enceladus day is 1.3 Earth days) are shown above each orbit level with corresponding color scheme in the operations schedule table. ``Ram" indicates that the instrument bus is pointed in the ramming direction relative to the plumes; for ``nadir" observations, instruments are pointed at Enceladus's surface. This observation schedule would allow THEO to conduct all the necessary science while complying with the solar power budget.
\label{obsschedule}%
}
\end{center}
\end{figure*}

To ensure timely completion of all science objectives and to reconcile our science objectives with the power and data constraints of operating under solar power at Saturn, THEO would have a specific observing schedule, summarized in Figure \ref{obsschedule}.  High-altitude orbits would be used to conduct the high-phase (backlit) plume imaging campaign from far enough away to catch the plumes at a sufficiently wide field of view. As it is also easier to do Enceladus orbit insertion at higher altitudes, we would begin our science mission at 500 km with 200 orbits of remote sensing with WAVES and DRIPS, the sub-mm instrument and the camera. This would be a unique set of observations in our schedule because the spacecraft would be oriented such that the instrument bus would be pointed in the ram direction. These observations would also be the only ones conducted with orbit nodes at midnight and noon to ensure proper lighting conditions for the high-phase images. \\

Such a configuration, however, is not sustainable power-wise; the solar panels would spend too much time in shadow.  In the subsequent observation campaigns, the spacecraft would be oriented such that its orbit nodes are at 6pm and 6am-- this modification is key to meeting the solar power requirements and comes at no cost to the science mission. The instrument bus would also point in the nadir direction. In this new orientation, THEO would conduct an additional 100 orbits with SWAMP, the mass spectrometer, at an altitude of 500 km to determine the plume composition at this distance from Enceladus's surface. \\

The spacecraft would then descend to an altitude of 100 km for only 10 orbits. This brief sampling time is enough to establish baseline measurements with SWAMP for what species make it to this lower altitude and for WAVES to make thermal emission maps of broader coverage to put the high resolution maps of later observations in the correct context.\\

The bulk of our observation campaign would be spent at 30 km, 18 km closer than the closest Cassini flyby of over Enceladus's south polar region. The highest priority science, making an inventory of the species within the plumes, would be conducted in the first 300 orbits.  WAVES would operate in the thermal mapping mode in high resolution strips. As the data collected by both instruments is relatively small in number of bits, SWAMP and WAVES would concurrently conduct their observations. On the next 300 orbits, DRIPS would map the south polar terrain, up to 50$^{\circ}$ S. OSMOSIS, the magnetometer suite, would operate for 10 orbits at 30 km and GEISER would conduct the Doppler-tracking experiment for 10 orbits at the 30 km altitude. \\

In the final stages of the proposed science mission, the spacecraft must exit Enceladus orbit to comply with planetary protection protocol \citep[e.g.][]{nasa2005planetary}. As the spacecraft climbs out of Enceladus orbit, SWAMP and WAVES would conduct an additional 20 orbits of observation at 100 km. The proposed fuel budget ensures enough fuel to de-orbit Enceladus and subsequently impact Tethys at the end of mission, thus eliminating any potential contamination of the THEO spacecraft crashing into Enceladus.\\

\subsection{Spacecraft Design}
\subsubsection{Mechanical and Configuration}
\begin{figure*}[h!]
\begin{center}
\includegraphics[width=1\textwidth]{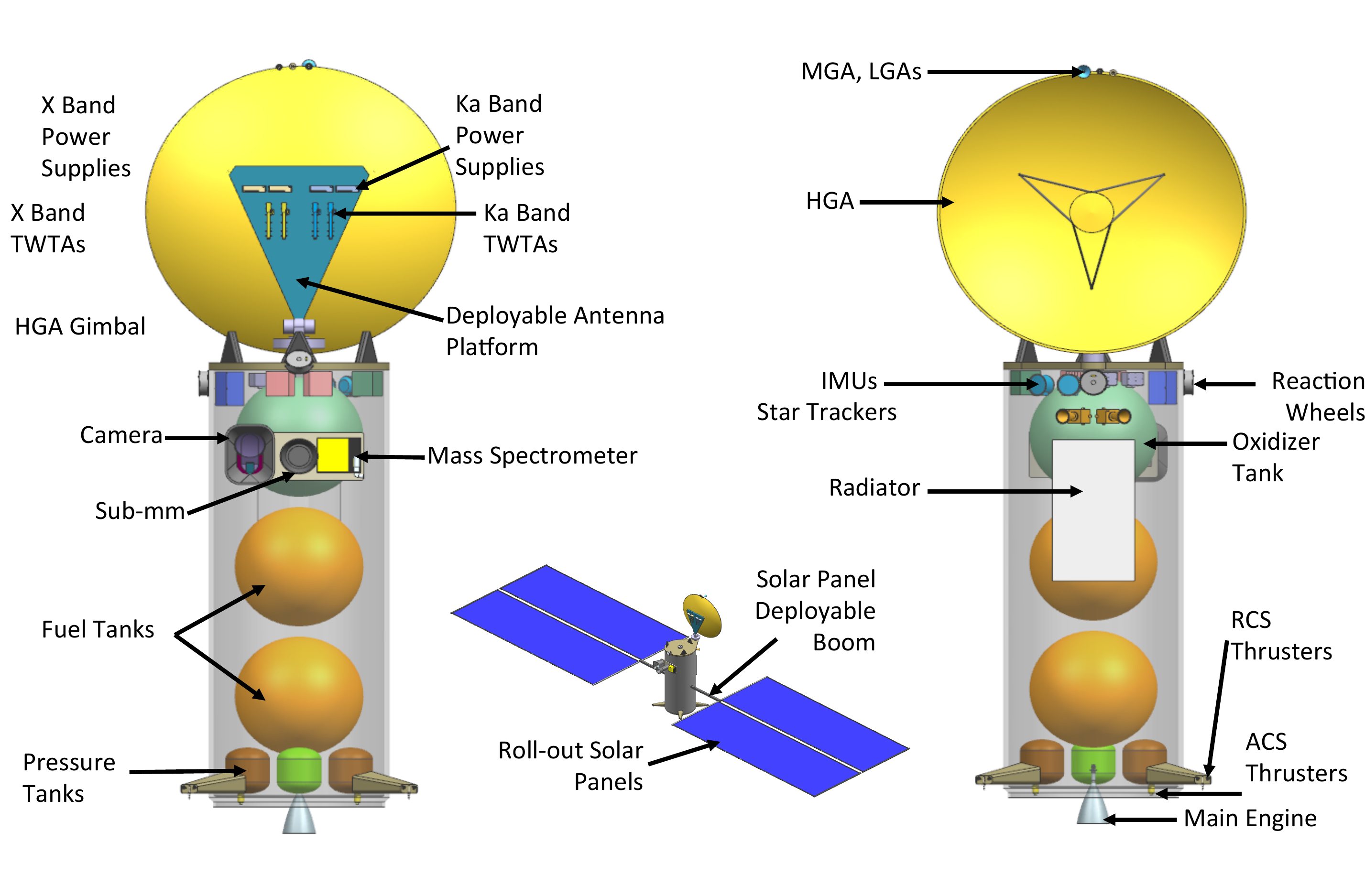}
\caption{CAD model for the THEO spacecraft. The solar arrays would roll out, an important characteristic for enabling the mission. Before deployment, the spacecraft is small enough to fit on an Atlas V for launch. Operationally, the higher packing density and larger panels would specifically enable the use of solar power at 10 AU.\label{spacecraft}}
\end{center}
\end{figure*}

The THEO spacecraft would be an orbiter in the family of JUNO and is shown in Figure \ref{spacecraft}. The main bus would be a standard cylindrical body of 4.5 m in height, 1.5 m in diameter, made of metal and metallic honeycomb. Dry mass of the proposed spacecraft is estimated at 1153.3 kg; wet mass is estimated at 4187 kg. As such, THEO would fit on the Atlas V 541 with the standard launch vehicle adapter and a mass margin of  2\%. A HGA would sit atop the cylinder and, as it is attached via a gimbal, have 2 degrees of motion. A medium gain antenna and two low gain antennae would be located at the top of the HGA.  The bus would hold three spherical tanks: one for the oxidizer and two for fuel. The main engine would be located opposite the HGA. The instrument suite with the mass spectrometer, sub-mm, and camera would be on the nadir-pointing side of the bus. To account for the relative velocity of the vapor, the mass spectrometer would be mounted at a 80$^{\circ}$ from nadir-pointing. \\

\subsubsection{Solar Panels}
Two wings of two Roll-Out Solar Array (ROSA) solar panels each would also attach to the main bus with 1 axis gimbals, allowing the flexible arrays to track the sunlight. Each wing would be made of two panels that are 36 m$^2$ in area. Once released from the rocket capsule, the solar arrays would be deployed in an accordion unfolding, similar to the deployment planned for NASA's Surface Water Ocean Topography mission. The cumulative estimated power output is 44.2 kW at beginning-of-life and 594 W at end-of-life operations at Saturn (a value consistent with the empirically-based predictions of \citet{lorenz2015energy} for the energy needed to transmit data back to Earth).\\

\subsubsection{Propulsion}
A dual-mode bi-propellant system with a 445 N main engine would provide spacecraft propulsion. Orbit changing maneuvers would be powered by N$_2$H$_4$ fuel and N$_2$O$_4$ oxidizer. N$_2$H$_4$ monopropellant would fuel the four 22 N RCS thrusters and twelve 0.9 N ACS thrusters. The proposed thrusters and propellant tanks are commercially available, minimizing development costs. 
The primary function of the RCS thrusters is to correct for misalignment in maneuvers executed by the main engine. It is assumed that the main engine thrust can only be aligned to within 1\%. Thus, the RCS thrusters are assigned approximately 23 m/s of $\Delta$v (1\% of the total $\Delta$v allocated to the main engine) to perform these corrections. These designs are flexible enough to accommodate mass growth up to the launch vehicle capability. \\ 

\subsubsection{Attitude Control}
THEO's proposed attitude control subsystem (ACS) components consist of commercial, off-the-shelf sensors and actuators whose operations would be within the required pointing, stability, and slewing margins. The 12 ACS thrusters provide contingency attitude control of up to 50 m/s of $\Delta$v for maneuvers such as momentum dumping and fast slewing. The sensor suite for determining position and pointing would include two Galileo AA-STAR star trackers, two Honeywell MIMU internal measurement units, and eight Adcole course grain sun sensors. Four RSI 4-33 reaction wheels would serve as THEO's actuators. THEO's predicted pointing ability is accurate to within 0.05$^{\circ}$, smaller than the 0.15$^{\circ}$ required. Pointing stability is predicted to be 1.7 arcsec/sec, a margin of 0.3 arcsec/sec. The predicted slewing velocity, important for our sub-mm plume observations at 500 km in the ram direction is three times greater than the minimum required. \\

\subsubsection{Thermal}
At Enceladus orbit, 10 AU, the spacecraft must maintain its own operational temperature (-40 $^{\circ}$C for the main bus). The spacecraft fuel, for example, must be warm enough to maintain its liquid phase and is therefore the driving factor in the thermal-related power requirements. Without the heat generated from a radioisotope thermoelectric generator power source, the main bus's temperature would be regulated by a suite of thermal hardware including  mechanical thermostats, platinum resistance thermometers, Kapton film resistive heaters, multilayer insulation, and louvers. While the temperature at the location where instruments attach to the spacecraft would be maintained by this system, instruments would provide their own independent thermal systems based on their appropriate operating temperatures. The spacecraft would also be covered with multi-layered insulation. \\

\subsubsection{Telecommunications}
THEO's communications operations related to spacecraft health and safety tracking as well as navigation would be sent using the X band and the high gain antenna. During all critical events, like Saturn Orbit Insertion and Enceladus Orbit Insertion, THEO would be in contact with operations control through continuous X band communications. \\

Power constraints and operational costs would prohibit the spacecraft from continuously sending back data. Therefore, the proposed spacecraft telecommunications operations includes storing and transmitting/receiving modes. To accommodate all data collected in our observational schedule, data compression would be used to limit data buildup on the spacecraft computer to 6 Gbits/ Earth day. The computer hardware would consist of a dual-string Rad-750 with 786 Gbits of mass storage with redundant 1553 buses for interfacing the main computer and other systems. During downlink with the 34 m arrays of the Deep Space Network, the high gain antenna would send these data at 40 kbps via the Ka band (7.5 mm - 1 cm) for four hours each day. Thus, THEO would send back ~0.5 Gbits per day during data collecting modes.  Upon receipt and timely validation, the data would be archived on the Planetary Data System.\\

\subsection{Key Trades}
\label{trades}
Using solar power instead of a radioisotope thermoelectric generator (RTG) was a key decision in our concept study. As is discussed in Section \ref{risk}, solar power has not been demonstrated at 10 AU and thus represents some level of risk. However, the demonstrated ability and power output of an RTG system did not balance the mass and financial costs associated with it. Furthermore, RTGs did not specifically enable additional or better science. In terms of solar power architecture, the most challenging science tasks are those involving eclipse geometries, the number and duration of which limit solar charging capabilities. We designed THEO to easily accommodate these difficulties with the observing schedule summarized in Figure \ref{obsschedule}. With a little planning, therefore, THEO would operate within the power constraints of a solar powered spacecraft without sacrificing any of the science tasks outlined in Figure \ref{STM}. \\

Another key trade explored in the development of THEO's mission architecture was the choice between a flyby mission or an orbiter. The $\Delta$v difference for achieving orbit versus a multiple flyby configuration was on the order of a few 100 m/s (less than 10\% of the mission total) and as the mission design was largely the same, the difference in operations cost was also minimal. The difference in science capabilities, however, was significant. Twelve flybys simply could not gather enough data to meet the science goals outlined in Section 2, whereas the 930 orbits of data collection over three altitudes of an orbiter configuration could. The biggest disadvantage to an orbiter configuration is the slower orbital velocity-- it is insufficient to pulverize ice grains with kinetic energy alone. However, as discussed above, lower velocities present advantages to measuring the plume vapor (such as preserving larger molecules). Therefore, we chose to conduct our investigation with the mass spectrometer as designed, meeting the science objectives listed in Section \ref{goals}. Future investigations could address the possibility of adding a pulverizing mechanism to the mass spectrometer.\\

\section{Management}
\label{management}
The THEO mission would be a ``standard" New Frontiers mission. The spacecraft, as designed, incorporates only technology at high technology readiness level, minimizing obstacles in technology development. THEO's development timeline would therefore be consistent with other New Frontiers missions. The spacecraft design is also of low enough mass to take advantage of launch vehicle incentives, which provide a cost credit that allows THEO to fit comfortably within the New Frontiers cost cap. Thus, the assessment of Enceladus's habitability is possible with modest investment and minimal risk.\\

\subsection{Cost}
\label{cost}
The nominal 2009 New Frontiers mission cost cap is set at \$1000 M (excluding the cost of the launch vehicle). The THEO spacecraft, however, is sufficiently light to fit within a smaller launch vehicle, the Atlas V 541. This would activate a cost incentive and raise the cost cap to \$1046.9 M. In Table \ref{costtable}, the current and predicted best estimate costs for each development phase are listed. The current best estimate includes project, development, and operating costs, excluding the cost of the launch vehicle. The predicted best estimate is calculated by adding reserves to the current best estimate. These figures were derived using Team-X Institutional (JPL) cost models. The total proposed mission cost is estimated to be \$1011.1 M, 3.4\% below the cap. \\

\begin{table*} 
    \begin{tabular}{ l c c c }
         & Current Best Estimate & Reserve & Predicted Best Estimate\\
          &(\$M)&&(\$M)\\ 
         \hline
        Phase A & 2.1  & 20\%  & 2.5 \\ 
        Phase B & 54.2  & 30\% & 70.4 \\ 
        Phase C-D & 541.9 & 30  & 704.2  \\ 
        Operations Cost & 206.2 & 13 & 233.9 \\
        (Phases E-F)& &  &     
    \end{tabular} 
    \caption{Total cost estimates for the concept mission. Current best estimates are calculated without reserves; predicted best estimates add reserves to the current best estimate. Dollar amounts are listed in millions for each phase of the mission, from development through operations.\label{costtable}}
\end{table*}

There are no reductions in mission scope (instrument payload or mission schedule) that would result in worthwhile savings in cost, mass, power, data, or time without significantly undermining the science mission. Thus, the threshold mission is the baseline mission for THEO. There is no descope-related cost model for the mission. \\

During the developmental phases (A-D), the payload and flight systems represent the largest portions of the \$777.1 M total at 14\% and 38\% respectively.  33\% of the flight systems cost is budgeted for the power system, including the manufacturing and installation related to the solar panels. \$178.9 M (23\%) is retained as development reserves. Operations costs (E-F) are estimated to total \$233.9 M. As mission operations-- DSN tracking; personnel for monitoring, navigation, and mission planning; etc. -- represents the largest portion of these phases's cost at 54\%, we budget an additional 13\% for phases E-F as reserves.\\

\subsection{Risk}
\label{risk}

Risk assessment was conducted in association with Team X using the risk assessment matrix per subsystem outlined in \citet{hihn2010identification}. The most significant risks identified in our study reflect the novelty of our proposed mission design. THEO would utilize solar panels and a mass spectrometer whose heritages are, at the time of mission design, not proven on a successfully flown mission. (That's not to say that solar power hasn't been proposed for Saturn before-- see, for example, the Enceladus Life Finder mission concept of \citet{2015LPI....46.1525L}.)  In addition to these technological risks, we find two low level risks associated with the proposed mission: the thermal environment of the inner solar system trajectory and the reliability of plume activity at Enceladus. \\ 

We consider the primary risk in this mission concept to be the utilization of solar cells at Saturn-- i.e., at an unprecedented distance from the Sun. This risk is accepted in lieu of the programmatic risks of nuclear power systems as the mass savings of solar power enables the scope of our science mission. Several mitigation strategies are built into the mission concept. The array design incorporates a 43\% margin to accommodate array degradation or unforeseen operational constraints, and meets spacecraft power requirements at solar incidence angles of 10$^{\circ}$ (though gimbaled arrays will track the sunlight, providing significantly higher levels of insolation). THEO utilizes ROSA technology currently being developed under contract to NASA's Space Technology Mission Directorate.  Missions currently in development (e.g. the Asteroid Redirect Robotic Mission) are expected to advance the flight-readiness of the hardware and establish test and fabrication practices that would be used by THEO. Though engineering and operational challenges can be overcome, the THEO mission cannot be flown as a New Frontiers class mission with the science payload slated without ROSA technology. Thus, dependence on this not-yet-flown array poses a critical risk to the THEO mission concept.\\

The readiness level of the mass spectrometer proposed for THEO depends upon successful development of the MASPEX instrument currently slated to fly aboard NASA's Europa mission. Should MASPEX not be successfully flown prior to THEO development, THEO's mass spectrometer could instead be modeled after ROSINA, the mass spectrometer onboard Rosetta. The ROSINA instrument is smaller, less expensive, and uses less power than MASPEX. However, ROSINA is not capable of resolving nucleotides and other organic molecules $>$300 amu \citep{balsiger2007rosina}. There is thus a science cost to this strategy: we can establish habitability, but not life. Because there are a number of important science tasks still possible with a ROSINA-equivalent model, we consider the science loss an acceptable one and therefore designate the risk to THEO as low.\\ 

The Earth-Venus-Earth-Earth trajectory proposed for THEO incurs significant variations in thermal environment as the spacecraft initially travels in closer to the Sun for a Venus gravity assist. The risk to the spacecraft would be mitigated operationally by using the high-gain antenna as a sun shield, a technique successfully used by the Cassini spacecraft \citep{matson2003cassini}.\\

As our proposed mission relies on the \emph{in situ} sampling opportunity offered by Enceladus's plumes, we must address the risk of the plumes not being active. The likelihood of sustained activity is high based on several lines of evidence. Plume activity has persisted throughout the \emph{Cassini} mission and has been shown to be the source of Saturn's E-ring, consistently observed since its discovery in 1967 \citep{1967Natur.214..793F}. (Particle lifetimes in the E-ring are estimated at most to be $\sim$50 years due to sputtering from energetic particles in the Saturnian system \citep{2001Icar..149..384J}.) Additionally, water in the magnetosphere \citep{2008Natur.456..477H} constrains activity to at least within the last 15 years. Tectonic features away from Enceladus's currently active south polar region suggest that venting of the subsurface ocean may have occurred throughout Enceladus's history \citep{2016Icar..264...37T}. Therefore, the plumes are not expected to cease within the geologically near future. \\

However, in the worst-case scenario that upon Saturn system arrival Enceladus is no longer active, THEO could still achieve meaningful science. The THEO instrument package is capable of alternative scientific investigations in the Saturn system which could be implemented in the event that Earth-based observations revealed cessation of plume activity during THEO's cruise. Detailed descriptions of these alternatives are beyond the scope of this work. The spacecraft $\Delta$v budget would be sufficient to modify the trajectory and enter Saturn-orbit, or to take advantage of the numerous flybys that would otherwise be used to pump-down the vehicle's velocity before entering Enceladus orbit. Alternatively, exploring a newly inactive Enceladus might also provide valuable insight into the mechanics of plume activity and shutoff. Though the resulting science investigation would be drastically different from that proposed here, we emphasize that the proposed instrument package is inherently flexible; the spacecraft would not ``go to waste" in the unlikely event that the plumes mysteriously cease.\\

\section{Conclusions}
\label{conclusions}
As \emph{Cassini} approaches its end of mission in 2017, it is increasingly important that the community consider the next phase of exploring the dynamic worlds revealed by the Saturnian flagship. Enceladus is one such world-- its plumes offer a unique opportunity to sample a potentially habitable subsurface ocean with relative ease. Other proposed missions have recognized the astrobiological potential of Enceladus, including sample return \citep{2012AsBio..12..730T}, joint Titan-Enceladus investigations \citep{2009ExA....23..893C,2011LPI....42.1326S,2014P&SS..104...59T}, and plume-sampling on a Discovery-class budget \citep{2015LPI....46.1525L}. Like these other mission concepts, THEO was specifically designed for exploring Enceladus to answer questions uncovered by \emph{Cassini} discoveries. \\

\begin{figure}[h!]
\begin{center}
\includegraphics[width=0.7\columnwidth]{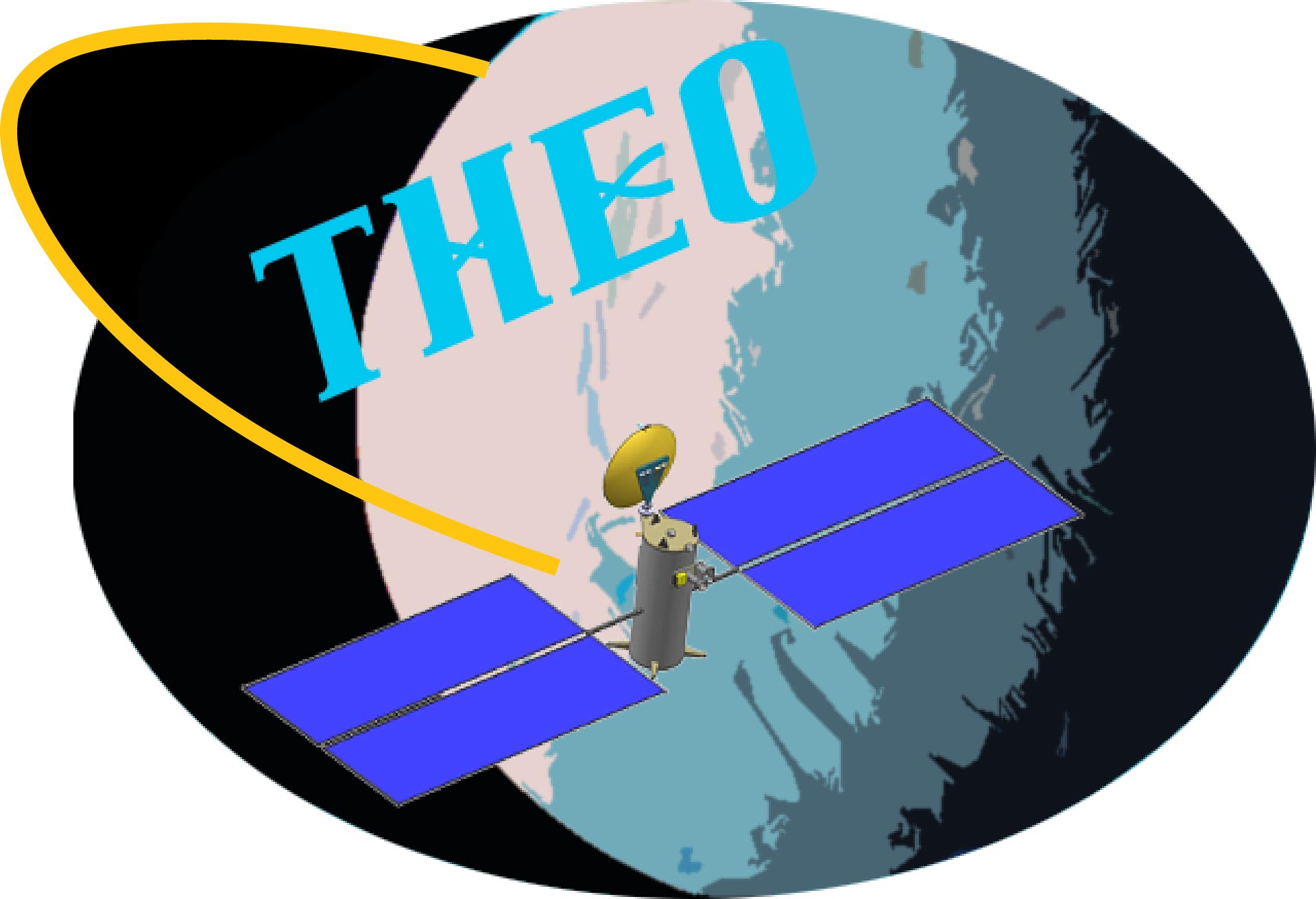}
\caption{Mission logo for Testing the Habitability of Enceladus's Ocean. The proposed patch highlights the three key characteristics of THEO: solar-paneled spacecraft (featured center), in orbit around Enceladus (yellow trajectory), seeking to answer questions related to habitability and therefore biology (double strands in the acronym lettering).
\label{logo}%
}
\end{center}
\end{figure}

THEO would meet a preponderance of Enceladus science goals laid out by the 2013 Decadal Survey by conducting experiments that would elucidate both whether the moon's hidden ocean is habitable and the factors that affect that answer. We think that these two questions-- whether life exists somewhere and why it does or does not-- are both necessary questions when seeking life in the solar system and are thus the driving motivation for THEO. The proposed science mission (the logo for which is shown in Figure \ref{logo}) would address how Enceladus's plumes are connected to its subsurface ocean, what mechanisms might be keeping the ocean liquid, whether the abiotic conditions of the ocean are suitable for Earth-analogue life forms, and whether there is evidence of biological processes active at the time of sampling. The THEO mission concept demonstrates that a medium-class mission of a solar-powered orbiter can take full advantage of Enceladus's plumes to explore a potentially habitable ocean world of the outer solar system. \\

\section{Acknowledgements}
Disclaimer: The cost information contained in this document is of a budgetary and planning nature and is intended for informational purposes only.  It does not constitute a commitment on the part of JPL and/or Caltech.\\

This work was carried out at JPL/Caltech under a contract with NASA. \\

We thank Chris McKay and an anonymous reviewer for their helpful comments as well as Hunter Waite for his generous insight into the MASPEX instrument.\\

The THEO team would like to thank the Planetary Science Summer School mentoring team and staff whose efforts made the experience possible: Anita M. Sohus, Leslie L. Lowes, Jessica Parker, and Greg Baerg. We also gratefully acknowledge the TeamX members for their patience, expertise, and insight: Pamela Clark, Alfred Nash, Timothy Koch, Matt Bennett, Austin  Lee, Roger Klemm, Enrique Baez, Brian Bairstow, Adam Nelessen, Gregory Welz, Melissa Vick, Morgan Henry, Ted Sweetser, Ron Hall, Jarius Hihn, Masashi Mizukami, Bill Smythe, Michael Sekerak, Vaughn Cable, Eric  Sunada, Daniel Forgette, Jennifer Miller, Yogi  Krikorian, Try Lam, Dhack Muthulingham, and Patrick Ward. As the culminating event of our summer school, the following volunteers served on a review panel: Farah Alibay, Luther Beegle, Richard Bennett, Ramon P. DePaula, Kevin Hand, Keith Grogan, Young Lee, and Rob Sherwood. We thank the panel for their time and constructive critiques that strengthened our  ``proposal" and thus the results presented in this paper. Thanks also to task managers J. Cutts, M. Viotti, and R. Zimmerman-Brachman. Finally, we express our gratitude to NASA HQ Science Mission Directorate, the NASA Planetary Science Division, and the NASA Radioisotope Power Systems Program for providing continued financial support for JPL's Planetary Science Summer school. \\
SMM acknowledges support from the NASA Earth and Space Science Fellowship Program- Grant NNX14AO30H.\\

\subsection{Author Contributions}
SMM led the mission as Principal Investigator and wrote the manuscript. 
TEC served as Project Manager and played a primary role in writing the manuscript. 
CMPL served as the risk and programmatics chair, geochemistry team lead, and provided revisions to the manuscript. 
ENS was the systems engineer, co-lead for the sub-mm instrument, and assisted with manuscript reorganization and revisions.
JDH was the Deputy PI, geology team lead, telecommunications subsystems chair, co-lead on the imaging camera, and assisted with manuscript revisions. 
VZS served as Deputy System Engineer, co-lead on the imaging camera, and assisted with manuscript revisions. 
KEP led the command and data handling subsystem and assisted with manuscript revisions. 
CJS served as the attitude control subsystem chair, the magnetometer suite lead, and provided revisions to the manuscript. 
JGO was as the science chair, led the geophysics team, and assisted with manuscript revisions. 
JKD served as the ground systems chair, lead of the Plumes team, and co-lead for the mass spectrometer.
CWSL served as the mission design chair, led the sub-mm instrument, and assisted with manuscript revisions. 
EMP was in charge of propulsion chair and assisted with manuscript revisions. 
JJW led the thermal control subsystem and provided revisions to the manuscript. 
SP was in charge of the power design, served as co-lead for the mass spectrometer, and assisted with manuscript revisions. 
MC served as the instruments lead and assisted with manuscript revisions.
KKJ was in charge of spacecraft configuration and assisted with manuscript revisions.
AK served as the cost chair and assisted with manuscript revisions.
KD was in charge of mechanical design.
CJB and KLM were the coordinators of the JPL Planetary Science Summer School.
\bibliographystyle{elsarticle-harv.bst}

\bibliography{fullarticle.bbl}

\begin{thebibliography}{106}
\expandafter\ifx\csname natexlab\endcsname\relax\def\natexlab#1{#1}\fi
\expandafter\ifx\csname url\endcsname\relax
  \def\url#1{\texttt{#1}}\fi
\expandafter\ifx\csname urlprefix\endcsname\relax\def\urlprefix{URL }\fi

\bibitem[{Allen et~al.(2006)Allen, Sherwood~Lollar, Runnegar, Oehler, Lyons,
  Manning, and Summers}]{allen2006mars}
Allen, M., Sherwood~Lollar, B., Runnegar, B., Oehler, D.~Z., Lyons, J.~R.,
  Manning, C.~E., Summers, M.~E., 2006. {Is Mars Alive?} Eos, Transactions
  American Geophysical Union 87~(41), 433--439.

\bibitem[{Baland et~al.(2015)Baland, Yseboodt, and
  Van~Hoolst}]{baland2015obliquity}
Baland, R.-M., Yseboodt, M., Van~Hoolst, T., 2015. {The obliquity of
  Enceladus}. Icarus.

\bibitem[{{Balsiger} et~al.(2007){Balsiger}, {Altwegg}, {Bochsler},
  et~al.}]{balsiger2007rosina}
{Balsiger}, H., {Altwegg}, K., {Bochsler}, P., et~al., 2007. {Rosina--Rosetta
  orbiter spectrometer for ion and neutral analysis}. Space Science Reviews
  128~(1-4), 745--801.

\bibitem[{{Barr} and {Preuss}(2010)}]{2010Icar..208..499B}
{Barr}, A.~C., {Preuss}, L.~J., jul 2010. {On the origin of south polar folds
  on Enceladus}. Icarus 208, 499--503.

\bibitem[{{Baum} et~al.(1981){Baum}, {Kreidl}, {Westphal},
  et~al.}]{1981Icar...47...84B}
{Baum}, W.~A., {Kreidl}, T., {Westphal}, J.~A., et~al., jul 1981. {Saturn's E
  ring}. Icarus 47, 84--96.

\bibitem[{{Bland} et~al.(2012){Bland}, {Singer}, {McKinnon}, and
  {Schenk}}]{2012GeoRL..3917204B}
{Bland}, M.~T., {Singer}, K.~N., {McKinnon}, W.~B., {Schenk}, P.~M., sep 2012.
  {Enceladus' extreme heat flux as revealed by its relaxed craters}.
  Geophysical Research Letters 39, 17204.

\bibitem[{{Bouquet} et~al.(2015){Bouquet}, {Mousis}, {Waite}, and
  {Picaud}}]{2015GeoRL..42.1334B}
{Bouquet}, A., {Mousis}, O., {Waite}, J.~H., {Picaud}, S., mar 2015. {Possible
  evidence for a methane source in Enceladus' ocean}. Geophysical Research
  Letters 42, 1334--1339.

\bibitem[{{Brown} et~al.(2006){Brown}, {Clark}, {Buratti}, {Cruikshank},
  {Barnes}, {Mastrapa}, {Bauer}, {Newman}, {Momary}, {Baines}, {Bellucci},
  {Capaccioni}, {Cerroni}, {Combes}, {Coradini}, {Drossart}, {Formisano},
  {Jaumann}, {Langevin}, {Matson}, {McCord}, {Nelson}, {Nicholson}, {Sicardy},
  and {Sotin}}]{2006Sci...311.1425B}
{Brown}, R.~H., {Clark}, R.~N., {Buratti}, B.~J., {Cruikshank}, D.~P.,
  {Barnes}, J.~W., {Mastrapa}, R.~M.~E., {Bauer}, J., {Newman}, S., {Momary},
  T., {Baines}, K.~H., {Bellucci}, G., {Capaccioni}, F., {Cerroni}, P.,
  {Combes}, M., {Coradini}, A., {Drossart}, P., {Formisano}, V., {Jaumann}, R.,
  {Langevin}, Y., {Matson}, D.~L., {McCord}, T.~B., {Nelson}, R.~M.,
  {Nicholson}, P.~D., {Sicardy}, B., {Sotin}, C., Mar. 2006. {Composition and
  Physical Properties of Enceladus' Surface}. Science 311, 1425--1428.

\bibitem[{{B{\v e}hounkov{\'a}} et~al.(2010){B{\v e}hounkov{\'a}}, {Tobie},
  {Choblet}, and {{\v C}adek}}]{2010JGRE..115.9011B}
{B{\v e}hounkov{\'a}}, M., {Tobie}, G., {Choblet}, G., {{\v C}adek}, O., sep
  2010. {Coupling mantle convection and tidal dissipation: Applications to
  Enceladus and Earth-like planets}. Journal of Geophysical Research (Planets)
  115, 9011.

\bibitem[{{B{\v e}hounkov{\'a}} et~al.(2012){B{\v e}hounkov{\'a}}, {Tobie},
  {Choblet}, and {{\v C}adek}}]{2012Icar..219..655B}
{B{\v e}hounkov{\'a}}, M., {Tobie}, G., {Choblet}, G., {{\v C}adek}, O., jun
  2012. {Tidally-induced melting events as the origin of south-pole activity on
  Enceladus}. Icarus 219, 655--664.

\bibitem[{{B{\v e}hounkov{\'a}} et~al.(2015){B{\v e}hounkov{\'a}}, {Tobie},
  {{\v C}adek}, {Choblet}, {Porco}, and {Nimmo}}]{2015NatGe...8..601B}
{B{\v e}hounkov{\'a}}, M., {Tobie}, G., {{\v C}adek}, O., {Choblet}, G.,
  {Porco}, C., {Nimmo}, F., aug 2015. {Timing of water plume eruptions on
  Enceladus explained by interior viscosity structure}. Nature Geoscience 8,
  601--604.

\bibitem[{{Campagnola} et~al.(2010){Campagnola}, {Strange}, and
  {Russell}}]{2010CeMDA.108..165C}
{Campagnola}, S., {Strange}, N.~J., {Russell}, R.~P., Oct. 2010. {A fast tour
  design method using non-tangent v-infinity leveraging transfer}. Celestial
  Mechanics and Dynamical Astronomy 108, 165--186.

\bibitem[{{Coustenis} et~al.(2009){Coustenis}, {Atreya}, {Balint}, {Brown},
  {Dougherty}, {Ferri}, {Fulchignoni}, {Gautier}, {Gowen}, {Griffith},
  {Gurvits}, {Jaumann}, {Langevin}, {Leese}, {Lunine}, {McKay}, {Moussas},
  {M{\"u}ller-Wodarg}, {Neubauer}, {Owen}, {Raulin}, {Sittler}, {Sohl},
  {Sotin}, {Tobie}, {Tokano}, {Turtle}, {Wahlund}, {Waite}, {Baines},
  {Blamont}, {Coates}, {Dandouras}, {Krimigis}, {Lellouch}, {Lorenz}, {Morse},
  {Porco}, {Hirtzig}, {Saur}, {Spilker}, {Zarnecki}, {Choi}, {Achilleos},
  {Amils}, {Annan}, {Atkinson}, {B{\'e}nilan}, {Bertucci}, {B{\'e}zard},
  {Bjoraker}, {Blanc}, {Boireau}, {Bouman}, {Cabane}, {Capria},
  {Chassefi{\`e}re}, {Coll}, {Combes}, {Cooper}, {Coradini}, {Crary},
  {Cravens}, {Daglis}, {de Angelis}, {de Bergh}, {de Pater}, {Dunford},
  {Durry}, {Dutuit}, {Fairbrother}, {Flasar}, {Fortes}, {Frampton}, {Fujimoto},
  {Galand}, {Grasset}, {Grott}, {Haltigin}, {Herique}, {Hersant}, {Hussmann},
  {Ip}, {Johnson}, {Kallio}, {Kempf}, {Knapmeyer}, {Kofman}, {Koop}, {Kostiuk},
  {Krupp}, {K{\"u}ppers}, {Lammer}, {Lara}, {Lavvas}, {Le Mou{\'e}lic},
  {Lebonnois}, {Ledvina}, {Li}, {Livengood}, {Lopes}, {Lopez-Moreno}, {Luz},
  {Mahaffy}, {Mall}, {Martinez-Frias}, {Marty}, {McCord}, {Menor Salvan},
  {Milillo}, {Mitchell}, {Modolo}, {Mousis}, {Nakamura}, {Neish}, {Nixon}, {Nna
  Mvondo}, {Orton}, {Paetzold}, {Pitman}, {Pogrebenko}, {Pollard},
  {Prieto-Ballesteros}, {Rannou}, {Reh}, {Richter}, {Robb}, {Rodrigo},
  {Rodriguez}, {Romani}, {Ruiz Bermejo}, {Sarris}, {Schenk}, {Schmitt},
  {Schmitz}, {Schulze-Makuch}, {Schwingenschuh}, {Selig}, {Sicardy},
  {Soderblom}, {Spilker}, {Stam}, {Steele}, {Stephan}, {Strobel}, {Szego},
  {Szopa}, {Thissen}, {Tomasko}, {Toublanc}, {Vali}, {Vardavas}, {Vuitton},
  {West}, {Yelle}, and {Young}}]{2009ExA....23..893C}
{Coustenis}, A., {Atreya}, S.~K., {Balint}, T., {Brown}, R.~H., {Dougherty},
  M.~K., {Ferri}, F., {Fulchignoni}, M., {Gautier}, D., {Gowen}, R.~A.,
  {Griffith}, C.~A., {Gurvits}, L.~I., {Jaumann}, R., {Langevin}, Y., {Leese},
  M.~R., {Lunine}, J.~I., {McKay}, C.~P., {Moussas}, X., {M{\"u}ller-Wodarg},
  I., {Neubauer}, F., {Owen}, T.~C., {Raulin}, F., {Sittler}, E.~C., {Sohl},
  F., {Sotin}, C., {Tobie}, G., {Tokano}, T., {Turtle}, E.~P., {Wahlund},
  J.-E., {Waite}, J.~H., {Baines}, K.~H., {Blamont}, J., {Coates}, A.~J.,
  {Dandouras}, I., {Krimigis}, T., {Lellouch}, E., {Lorenz}, R.~D., {Morse},
  A., {Porco}, C.~C., {Hirtzig}, M., {Saur}, J., {Spilker}, T., {Zarnecki},
  J.~C., {Choi}, E., {Achilleos}, N., {Amils}, R., {Annan}, P., {Atkinson},
  D.~H., {B{\'e}nilan}, Y., {Bertucci}, C., {B{\'e}zard}, B., {Bjoraker},
  G.~L., {Blanc}, M., {Boireau}, L., {Bouman}, J., {Cabane}, M., {Capria},
  M.~T., {Chassefi{\`e}re}, E., {Coll}, P., {Combes}, M., {Cooper}, J.~F.,
  {Coradini}, A., {Crary}, F., {Cravens}, T., {Daglis}, I.~A., {de Angelis},
  E., {de Bergh}, C., {de Pater}, I., {Dunford}, C., {Durry}, G., {Dutuit}, O.,
  {Fairbrother}, D., {Flasar}, F.~M., {Fortes}, A.~D., {Frampton}, R.,
  {Fujimoto}, M., {Galand}, M., {Grasset}, O., {Grott}, M., {Haltigin}, T.,
  {Herique}, A., {Hersant}, F., {Hussmann}, H., {Ip}, W., {Johnson}, R.,
  {Kallio}, E., {Kempf}, S., {Knapmeyer}, M., {Kofman}, W., {Koop}, R.,
  {Kostiuk}, T., {Krupp}, N., {K{\"u}ppers}, M., {Lammer}, H., {Lara}, L.-M.,
  {Lavvas}, P., {Le Mou{\'e}lic}, S., {Lebonnois}, S., {Ledvina}, S., {Li}, J.,
  {Livengood}, T.~A., {Lopes}, R.~M., {Lopez-Moreno}, J.-J., {Luz}, D.,
  {Mahaffy}, P.~R., {Mall}, U., {Martinez-Frias}, J., {Marty}, B., {McCord},
  T., {Menor Salvan}, C., {Milillo}, A., {Mitchell}, D.~G., {Modolo}, R.,
  {Mousis}, O., {Nakamura}, M., {Neish}, C.~D., {Nixon}, C.~A., {Nna Mvondo},
  D., {Orton}, G., {Paetzold}, M., {Pitman}, J., {Pogrebenko}, S., {Pollard},
  W., {Prieto-Ballesteros}, O., {Rannou}, P., {Reh}, K., {Richter}, L., {Robb},
  F.~T., {Rodrigo}, R., {Rodriguez}, S., {Romani}, P., {Ruiz Bermejo}, M.,
  {Sarris}, E.~T., {Schenk}, P., {Schmitt}, B., {Schmitz}, N.,
  {Schulze-Makuch}, D., {Schwingenschuh}, K., {Selig}, A., {Sicardy}, B.,
  {Soderblom}, L., {Spilker}, L.~J., {Stam}, D., {Steele}, A., {Stephan}, K.,
  {Strobel}, D.~F., {Szego}, K., {Szopa}, C., {Thissen}, R., {Tomasko}, M.~G.,
  {Toublanc}, D., {Vali}, H., {Vardavas}, I., {Vuitton}, V., {West}, R.~A.,
  {Yelle}, R., {Young}, E.~F., mar 2009. {TandEM: Titan and Enceladus mission}.
  Experimental Astronomy 23, 893--946.

\bibitem[{{Cronin} and {Pizzarello}(1986)}]{cronin1986amino}
{Cronin}, J.~R., {Pizzarello}, S., 1986. {Amino acids of the Murchison
  meteorite. III. Seven carbon acyclic primary $\alpha$-amino alkanoic acids1}.
  Geochimica et cosmochimica acta 50~(11), 2419--2427.

\bibitem[{{Dougherty} et~al.(2006){Dougherty}, {Khurana}, {Neubauer},
  {Russell}, {Saur}, {Leisner}, and {Burton}}]{2006Sci...311.1406D}
{Dougherty}, M.~K., {Khurana}, K.~K., {Neubauer}, F.~M., {Russell}, C.~T.,
  {Saur}, J., {Leisner}, J.~S., {Burton}, M.~E., mar 2006. {Identification of a
  Dynamic Atmosphere at Enceladus with the Cassini Magnetometer}. Science 311,
  1406--1409.

\bibitem[{{Engel} et~al.(1994){Engel}, {Lunine}, and
  {Norton}}]{engel1994silicate}
{Engel}, S., {Lunine}, J.~I., {Norton}, D.~L., 1994. {Silicate interactions
  with ammonia-water fluids on early Titan}. Journal of Geophysical Research:
  Planets (1991--2012) 99~(E2), 3745--3752.

\bibitem[{{Feibelman}(1967)}]{1967Natur.214..793F}
{Feibelman}, W.~A., may 1967. {Concerning the ``D'' Ring of Saturn}. Nature
  214, 793--794.

\bibitem[{{Glein} et~al.(2015){Glein}, {Baross}, and
  {Waite}}]{2015GeCoA.162..202G}
{Glein}, C.~R., {Baross}, J.~A., {Waite}, J.~H., aug 2015. {The pH of
  Enceladus' ocean}. Geochimica et Cosmochimica Acta 162, 202--219.

\bibitem[{{Glein} et~al.(2016){Glein}, {Waite}, and
  {Lunine}}]{2016LPI....47.2885G}
{Glein}, C.~R., {Waite}, J.~H., {Lunine}, J.~I., Mar. 2016. {How Much
  Hydrothermal Hydrogen Might We Find in Enceladus' Plume?} In: Lunar and
  Planetary Science Conference. Vol.~47 of Lunar and Planetary Science
  Conference. p. 2885.

\bibitem[{{Goguen} et~al.(2013){Goguen}, {Buratti}, {Brown},
  et~al.}]{2013Icar..226.1128G}
{Goguen}, J.~D., {Buratti}, B.~J., {Brown}, R.~H., et~al., sep 2013. {The
  temperature and width of an active fissure on Enceladus measured with Cassini
  VIMS during the 14 April 2012 South Pole flyover}. Icarus 226, 1128--1137.

\bibitem[{{Gulkis} et~al.(2007){Gulkis}, {Frerking}, {Crovisier},
  et~al.}]{2007SSRv..128..561G}
{Gulkis}, S., {Frerking}, M., {Crovisier}, J., et~al., feb 2007. {MIRO:
  Microwave Instrument for Rosetta Orbiter}. Space Science Reviews 128,
  561--597.

\bibitem[{{Hansen} et~al.(2006){Hansen}, {Esposito}, {Stewart},
  et~al.}]{2006Sci...311.1422H}
{Hansen}, C.~J., {Esposito}, L., {Stewart}, A.~I.~F., et~al., mar 2006.
  {Enceladus' Water Vapor Plume}. Science 311, 1422--1425.

\bibitem[{{Hansen} et~al.(2008){Hansen}, {Esposito}, {Stewart},
  et~al.}]{2008Natur.456..477H}
{Hansen}, C.~J., {Esposito}, L.~W., {Stewart}, A.~I.~F., et~al., nov 2008.
  {Water vapour jets inside the plume of gas leaving Enceladus}. Nature 456,
  477--479.

\bibitem[{{Hansen} et~al.(2011){Hansen}, {Shemansky}, {Esposito},
  et~al.}]{2011GeoRL..3811202H}
{Hansen}, C.~J., {Shemansky}, D.~E., {Esposito}, L.~W., et~al., jun 2011. {The
  composition and structure of the Enceladus plume}. Geophysical Research
  Letters 38, 11202.

\bibitem[{{H{\"a}ssig} et~al.(2015){H{\"a}ssig}, {Libardoni}, {Mandt},
  {Miller}, and {Blase}}]{2015PSS..117..436H}
{H{\"a}ssig}, M., {Libardoni}, M., {Mandt}, K., {Miller}, G., {Blase}, R., nov
  2015. {Performance evaluation of a prototype multi-bounce time-of-flight mass
  spectrometer in linear mode and applications in space science}. Planetary and
  Space Science 117, 436--443.

\bibitem[{{Hedman} et~al.(2013){Hedman}, {Gosmeyer}, {Nicholson},
  et~al.}]{2013Natur.500..182H}
{Hedman}, M.~M., {Gosmeyer}, C.~M., {Nicholson}, P.~D., et~al., aug 2013. {An
  observed correlation between plume activity and tidal stresses on Enceladus}.
  Nature 500, 182--184.

\bibitem[{{Hedman} et~al.(2009){Hedman}, {Nicholson}, {Showalter}, {Brown},
  {Buratti}, and {Clark}}]{2009ApJ...693.1749H}
{Hedman}, M.~M., {Nicholson}, P.~D., {Showalter}, M.~R., {Brown}, R.~H.,
  {Buratti}, B.~J., {Clark}, R.~N., mar 2009. {Spectral Observations of the
  Enceladus Plume with Cassini-Vims}. ApJ 693, 1749--1762.

\bibitem[{{Hihn} et~al.(2010){Hihn}, {Chattopadhyay}, {Hanna}, {Port}, and
  {Eggleston}}]{hihn2010identification}
{Hihn}, J., {Chattopadhyay}, D., {Hanna}, R., {Port}, D., {Eggleston}, S.,
  2010. {Identification and classification of common risks in space science
  missions}. In: Proc. AIAA Space 2010 Conference and Exposition, Anaheim, CA.

\bibitem[{{Hill} and {Nuth}(2003)}]{hill2003catalytic}
{Hill}, H.~G.~M., {Nuth}, J.~A., 2003. {The catalytic potential of cosmic dust:
  implications for prebiotic chemistry in the solar nebula and other
  protoplanetary systems}. Astrobiology 3~(2), 291--304.

\bibitem[{{Hillier} et~al.(2007){Hillier}, {Green}, {McBride},
  et~al.}]{2007MNRAS.377.1588H}
{Hillier}, J.~K., {Green}, S.~F., {McBride}, N., et~al., jun 2007. {The
  composition of Saturn's E ring}. Monthly Notices of the Royal Astronomical
  Society 377, 1588--1596.

\bibitem[{{Hindermann} et~al.(1993){Hindermann}, {Hutchings}, and
  {Kiennemann}}]{hindermann1993mechanistic}
{Hindermann}, J.~P., {Hutchings}, G.~J., {Kiennemann}, A., 1993. {Mechanistic
  aspects of the formation of hydrocarbons and alcohols from CO hydrogenation}.
  Catalysis Reviews Science and Engineering 35~(1), 1--127.

\bibitem[{Horita and Berndt(1999)}]{horita1999abiogenic}
Horita, J., Berndt, M.~E., 1999. {Abiogenic methane formation and isotopic
  fractionation under hydrothermal conditions}. Science 285~(5430), 1055--1057.

\bibitem[{{Howett} et~al.(2011){Howett}, {Spencer}, {Pearl}, and
  {Segura}}]{2011JGRE..116.3003H}
{Howett}, C.~J.~A., {Spencer}, J.~R., {Pearl}, J., {Segura}, M., mar 2011.
  {High heat flow from Enceladus' south polar region measured using 10-600
  cm$^{-1}$ Cassini/CIRS data}. Journal of Geophysical Research (Planets) 116,
  3003.

\bibitem[{{Hsu} et~al.(2015){Hsu}, {Postberg}, {Sekine},
  et~al.}]{2015Natur.519..207H}
{Hsu}, H.-W., {Postberg}, F., {Sekine}, Y., et~al., mar 2015. {Ongoing
  hydrothermal activities within Enceladus}. Nature 519, 207--210.

\bibitem[{{Iess} et~al.(2014){Iess}, {Stevenson}, {Parisi}, {Hemingway},
  {Jacobson}, {Lunine}, {Nimmo}, {Armstrong}, {Asmar}, {Ducci}, and
  {Tortora}}]{2014Sci...344...78I}
{Iess}, L., {Stevenson}, D.~J., {Parisi}, M., {Hemingway}, D., {Jacobson},
  R.~A., {Lunine}, J.~I., {Nimmo}, F., {Armstrong}, J.~W., {Asmar}, S.~W.,
  {Ducci}, M., {Tortora}, P., apr 2014. {The Gravity Field and Interior
  Structure of Enceladus}. Science 344, 78--80.

\bibitem[{{Ingersoll} and {Ewald}(2011)}]{2011Icar..216..492I}
{Ingersoll}, A.~P., {Ewald}, S.~P., dec 2011. {Total particulate mass in
  Enceladus plumes and mass of Saturn's E ring inferred from Cassini ISS
  images}. Icarus 216, 492--506.

\bibitem[{{Ingersoll} and {Pankine}(2010)}]{2010Icar..206..594I}
{Ingersoll}, A.~P., {Pankine}, A.~A., apr 2010. {Subsurface heat transfer on
  Enceladus: Conditions under which melting occurs}. Icarus 206, 594--607.

\bibitem[{{Juh{\'a}sz} et~al.(2007){Juh{\'a}sz}, {Hor{\'a}nyi}, and
  {Morfill}}]{2007GeoRL..34.9104J}
{Juh{\'a}sz}, A., {Hor{\'a}nyi}, M., {Morfill}, G.~E., may 2007. {Signatures of
  Enceladus in Saturn's E ring}. Geophysical Research Letters 34, 9104.

\bibitem[{{Jurac} et~al.(2001){Jurac}, {Johnson}, and
  {Richardson}}]{2001Icar..149..384J}
{Jurac}, S., {Johnson}, R.~E., {Richardson}, J.~D., feb 2001. {Saturn's E Ring
  and Production of the Neutral Torus}. Icarus 149, 384--396.

\bibitem[{{Kamekura}(1998)}]{kamekura1998diversity}
{Kamekura}, M., 1998. {Diversity of extremely halophilic bacteria}.
  Extremophiles 2~(3), 289--295.

\bibitem[{Kelley et~al.(2001)Kelley, Karson, Blackman, Fru{\`E}h-Green,
  Butterfield, Lilley, Olson, Schrenk, Roe, Lebon, et~al.}]{kelley2001off}
Kelley, D.~S., Karson, J.~A., Blackman, D.~K., Fru{\`E}h-Green, G.~L.,
  Butterfield, D.~A., Lilley, M.~D., Olson, E.~J., Schrenk, M.~O., Roe, K.~K.,
  Lebon, G.~T., et~al., 2001. An off-axis hydrothermal vent field near the
  mid-atlantic ridge at 30 n. Nature 412~(6843), 145--149.

\bibitem[{{Kelley} et~al.(2005){Kelley}, {Karson}, {Fr{\"u}h-Green}, {Yoerger},
  {Shank}, {Butterfield}, {Hayes}, {Schrenk}, {Olson}, {Proskurowski},
  {Jakuba}, {Bradley}, {Larson}, {Ludwig}, {Glickson}, {Buckman}, {Bradley},
  {Brazelton}, {Roe}, {Elend}, {Delacour}, {Bernasconi}, {Lilley}, {Baross},
  {Summons}, and {Sylva}}]{2005Sci...307.1428K}
{Kelley}, D.~S., {Karson}, J.~A., {Fr{\"u}h-Green}, G.~L., {Yoerger}, D.~R.,
  {Shank}, T.~M., {Butterfield}, D.~A., {Hayes}, J.~M., {Schrenk}, M.~O.,
  {Olson}, E.~J., {Proskurowski}, G., {Jakuba}, M., {Bradley}, A., {Larson},
  B., {Ludwig}, K., {Glickson}, D., {Buckman}, K., {Bradley}, A.~S.,
  {Brazelton}, W.~J., {Roe}, K., {Elend}, M.~J., {Delacour}, A., {Bernasconi},
  S.~M., {Lilley}, M.~D., {Baross}, J.~A., {Summons}, R.~E., {Sylva}, S.~P.,
  Mar. 2005. {A Serpentinite-Hosted Ecosystem: The Lost City Hydrothermal
  Field}. Science 307, 1428--1434.

\bibitem[{{Kempf} et~al.(2008){Kempf}, {Beckmann}, {Moragas-Klostermeyer},
  {Postberg}, {Srama}, {Economou}, {Schmidt}, {Spahn}, and
  {Gr{\"u}n}}]{2008Icar..193..420K}
{Kempf}, S., {Beckmann}, U., {Moragas-Klostermeyer}, G., {Postberg}, F.,
  {Srama}, R., {Economou}, T., {Schmidt}, J., {Spahn}, F., {Gr{\"u}n}, E., feb
  2008. {The E ring in the vicinity of Enceladus. I. Spatial distribution and
  properties of the ring particles}. Icarus 193, 420--437.

\bibitem[{{Kempf} et~al.(2010){Kempf}, {Beckmann}, and
  {Schmidt}}]{2010Icar..206..446K}
{Kempf}, S., {Beckmann}, U., {Schmidt}, J., apr 2010. {How the Enceladus dust
  plume feeds Saturn's E ring}. Icarus 206, 446--457.

\bibitem[{{Kieffer} et~al.(2009){Kieffer}, {Lu}, {McFarquhar}, and
  {Wohletz}}]{2009Icar..203..238K}
{Kieffer}, S.~W., {Lu}, X., {McFarquhar}, G., {Wohletz}, K.~H., sep 2009. {A
  redetermination of the ice/vapor ratio of Enceladus'plumes: Implications for
  sublimation and the lack of a liquid water reservoir}. Icarus 203, 238--241.

\bibitem[{{Kivelson} et~al.(1997){Kivelson}, {Khurana}, {Joy}, {Russell},
  {Southwood}, {Walker}, and {Polanskey}}]{kivelson1997europa}
{Kivelson}, M.~G., {Khurana}, K.~K., {Joy}, S., {Russell}, C.~T., {Southwood},
  D.~J., {Walker}, R.~J., {Polanskey}, C., 1997. {Europa's magnetic signature:
  Report from Galileo's pass on 19 December 1996}. Science 276~(5316),
  1239--1241.

\bibitem[{{Kriegel} et~al.(2011){Kriegel}, {Simon}, {Motschmann}, {Saur},
  {Neubauer}, {Persoon}, {Dougherty}, and {Gurnett}}]{2011JGRA..11610223K}
{Kriegel}, H., {Simon}, S., {Motschmann}, U., {Saur}, J., {Neubauer}, F.~M.,
  {Persoon}, A.~M., {Dougherty}, M.~K., {Gurnett}, D.~A., oct 2011. {Influence
  of negatively charged plume grains on the structure of Enceladus' Alfv{\'e}n
  wings: Hybrid simulations versus Cassini Magnetometer data}. Journal of
  Geophysical Research (Space Physics) 116~(A15), A10223.

\bibitem[{{Lorenz}(2015)}]{lorenz2015energy}
{Lorenz}, R.~D., 2015. {Energy Cost of Acquiring and Transmitting Science Data
  on Deep-Space Missions}. Journal of Spacecraft and Rockets, 1--5.

\bibitem[{{Lunine} et~al.(2015){Lunine}, {Waite}, {Postberg}, {Spilker}, and
  {Clark}}]{2015LPI....46.1525L}
{Lunine}, J.~I., {Waite}, J.~H., {Postberg}, F., {Spilker}, L., {Clark}, K.,
  mar 2015. {Enceladus Life Finder: The Search for Life in a Habitable Moon}.
  In: Lunar and Planetary Science Conference. Vol.~46 of Lunar and Planetary
  Science Conference. p. 1525.

\bibitem[{{Marion} et~al.(2012){Marion}, {Kargel}, {Catling}, and
  {Lunine}}]{2012Icar..220..932M}
{Marion}, G.~M., {Kargel}, J.~S., {Catling}, D.~C., {Lunine}, J.~I., aug 2012.
  {Modeling ammonia-ammonium aqueous chemistries in the Solar System's icy
  bodies}. Icarus 220, 932--946.

\bibitem[{{Matson} et~al.(2007){Matson}, {Castillo}, {Lunine}, and
  {Johnson}}]{2007Icar..187..569M}
{Matson}, D.~L., {Castillo}, J.~C., {Lunine}, J., {Johnson}, T.~V., apr 2007.
  {Enceladus' plume: Compositional evidence for a hot interior}. Icarus 187,
  569--573.

\bibitem[{Matson et~al.(2003)Matson, Spilker, and Lebreton}]{matson2003cassini}
Matson, D.~L., Spilker, L.~J., Lebreton, J.-P., 2003. The cassini/huygens
  mission to the saturnian system. In: The Cassini-Huygens Mission. Springer,
  pp. 1--58.

\bibitem[{{McCollom}(1999)}]{mccollom1999methanogenesis}
{McCollom}, T.~M., 1999. {Methanogenesis as a potential source of chemical
  energy for primary biomass production by autotrophic organisms in
  hydrothermal systems on Europa}. Journal of Geophysical Research: Planets
  (1991--2012) 104~(E12), 30729--30742.

\bibitem[{{McCollom} and {Simoneit}(1999{\natexlab{a}})}]{mccollom1999abiotic}
{McCollom}, T.~M., {Simoneit}, B.~R.~T., 1999{\natexlab{a}}. {Abiotic formation
  of hydrocarbons and oxygenated compounds during thermal decomposition of iron
  oxalate}. Origins of Life and Evolution of the Biosphere 29~(2), 167--186.

\bibitem[{{McCollom} and {Simoneit}(1999{\natexlab{b}})}]{1999OLEB...29..167M}
{McCollom}, T.~M., {Simoneit}, B.~R.~T., mar 1999{\natexlab{b}}. {Abiotic
  Formation of Hydrocarbons and Oxygenated Compounds During Thermal
  Decomposition of Iron Oxalate}. Origins of Life and Evolution of the
  Biosphere 29, 167--186.

\bibitem[{McKay(2004)}]{mckay2004life}
McKay, C.~P., 2004. {What is life-and how do we search for it in other worlds?}
  PLoS Biology 2, 1260--1262.

\bibitem[{{McKay} et~al.(2014){McKay}, {Anbar}, {Porco}, and
  {Tsou}}]{2014AsBio..14..352M}
{McKay}, C.~P., {Anbar}, A.~D., {Porco}, C., {Tsou}, P., apr 2014. {Follow the
  Plume: The Habitability of Enceladus}. Astrobiology 14, 352--355.

\bibitem[{{McKay} et~al.(2008){McKay}, C., {Altheide}, {Davis}, and
  {Kral}}]{2008AsBio...8..909M}
{McKay}, C.~P., C., P.~C., {Altheide}, T., {Davis}, W.~L., {Kral}, T.~A., oct
  2008. {The Possible Origin and Persistence of Life on Enceladus and Detection
  of Biomarkers in the Plume}. Astrobiology 8, 909--919.

\bibitem[{{Miyakawa} et~al.(2002){Miyakawa}, {Cleaves}, and
  {Miller}}]{miyakawa2002cold}
{Miyakawa}, S., {Cleaves}, H.~J., {Miller}, S.~L., 2002. {The cold origin of
  life: A. Implications based on the hydrolytic stabilities of hydrogen cyanide
  and formamide}. Origins of Life and Evolution of the Biosphere 32~(3),
  195--208.

\bibitem[{{Mousis} et~al.(2009){Mousis}, {Lunine}, {Waite}, {Magee}, {Lewis},
  {Mandt}, {Marquer}, and {Cordier}}]{mousis2009formation}
{Mousis}, O., {Lunine}, J.~I., {Waite}, J.~H., {Magee}, B., {Lewis}, W.,
  {Mandt}, K.~E., {Marquer}, D., {Cordier}, D., 2009. {Formation conditions of
  Enceladus and origin of its methane reservoir}. The Astrophysical Journal
  Letters 701~(1), L39.

\bibitem[{{Nahm} and {Kattenhorn}(2015)}]{2015Icar..258...67N}
{Nahm}, A.~L., {Kattenhorn}, S.~A., sep 2015. {A unified nomenclature for
  tectonic structures on the surface of Enceladus}. Icarus 258, 67--81.

\bibitem[{NASA(2005)}]{nasa2005planetary}
NASA, 2005. {Planetary protection provisions for robotic extraterrestrial
  missions}.

\bibitem[{NASA(2009)}]{NFAO09}
NASA, 2009. {Announcement of Opportunity: New Frontiers 2009. NASA
  NNH09ZDA007O}.

\bibitem[{{National Research Council}(2011)}]{NAP13117}
{National Research Council}, 2011. {Vision and Voyages for Planetary Science in
  the Decade 2013-2022}.

\bibitem[{{Norton} and {Grant}(1988)}]{norton1988survival}
{Norton}, C.~F., {Grant}, W.~D., 1988. {Survival of halobacteria within fluid
  inclusions in salt crystals}. Journal of General Microbiology 134~(5),
  1365--1373.

\bibitem[{Parkinson et~al.(2008)Parkinson, Liang, Yung, and
  Kirschivnk}]{parkinson2008habitability}
Parkinson, C.~D., Liang, M.-C., Yung, Y.~L., Kirschivnk, J.~L., 2008.
  {Habitability of Enceladus: planetary conditions for life}. Origins of Life
  and Evolution of Biospheres 38~(4), 355--369.

\bibitem[{{Patthoff} and {Kattenhorn}(2011)}]{2011GeoRL..3818201P}
{Patthoff}, D.~A., {Kattenhorn}, S.~A., Sep. 2011. {A fracture history on
  Enceladus provides evidence for a global ocean}. Geophysical Research Letters
  38, L18201.

\bibitem[{{Perry} et~al.(2015){Perry}, {Teolis}, {Hurley},
  et~al.}]{2015Icar..257..139P}
{Perry}, M.~E., {Teolis}, B.~D., {Hurley}, D.~M., et~al., sep 2015. {Cassini
  INMS measurements of Enceladus plume density}. Icarus 257, 139--162.

\bibitem[{{Porco} et~al.(2014){Porco}, {DiNino}, and
  {Nimmo}}]{2014AJ....148...45P}
{Porco}, C., {DiNino}, D., {Nimmo}, F., sep 2014. {How the Geysers, Tidal
  Stresses, and Thermal Emission across the South Polar Terrain of Enceladus
  are Related}. The Astrophysical Journal 148, 45.

\bibitem[{{Porco} et~al.(2006){Porco}, {Helfenstein}, {Thomas},
  et~al.}]{2006Sci...311.1393P}
{Porco}, C.~C., {Helfenstein}, P., {Thomas}, P.~C., et~al., mar 2006. {Cassini
  Observes the Active South Pole of Enceladus}. Science 311, 1393--1401.

\bibitem[{Postberg(2015)}]{postberg2015refractory}
Postberg, F., 2015. Refractory organic compounds in enceladus? ice grains and
  hydrothermal activity. In: 2015 AGU Fall Meeting. Agu.

\bibitem[{{Postberg} et~al.(2009){Postberg}, {Kempf}, {Schmidt}, {Brilliantov},
  {Beinsen}, {Abel}, {Buck}, and {Srama}}]{2009Natur.459.1098P}
{Postberg}, F., {Kempf}, S., {Schmidt}, J., {Brilliantov}, N., {Beinsen}, A.,
  {Abel}, B., {Buck}, U., {Srama}, R., jun 2009. {Sodium salts in E-ring ice
  grains from an ocean below the surface of Enceladus}. Nature 459, 1098--1101.

\bibitem[{{Postberg} et~al.(2011){Postberg}, {Schmidt}, {Hillier},
  et~al.}]{2011Natur.474..620P}
{Postberg}, F., {Schmidt}, J., {Hillier}, J., et~al., jun 2011. {A salt-water
  reservoir as the source of a compositionally stratified plume on Enceladus}.
  Nature 474, 620--622.

\bibitem[{{Preston} and {Dartnell}(2014)}]{preston2014planetary}
{Preston}, L.~J., {Dartnell}, L.~R., 2014. {Planetary habitability: lessons
  learned from terrestrial analogues}. International Journal of Astrobiology
  13~(01), 81--98.

\bibitem[{Proskurowski et~al.(2008)Proskurowski, Lilley, Seewald,
  Fr{\"u}h-Green, Olson, Lupton, Sylva, and Kelley}]{proskurowski2008abiogenic}
Proskurowski, G., Lilley, M.~D., Seewald, J.~S., Fr{\"u}h-Green, G.~L., Olson,
  E.~J., Lupton, J.~E., Sylva, S.~P., Kelley, D.~S., 2008. {Abiogenic
  hydrocarbon production at Lost City hydrothermal field}. Science 319~(5863),
  604--607.

\bibitem[{{Rambaux} et~al.(2010){Rambaux}, {Castillo-Rogez}, {Williams}, and
  {Karatekin}}]{2010GeoRL..37.4202R}
{Rambaux}, N., {Castillo-Rogez}, J.~C., {Williams}, J.~G., {Karatekin}, {\"O}.,
  feb 2010. {Librational response of Enceladus}. Geophysical Research Letters
  37, 4202.

\bibitem[{{Roberts} and {Nimmo}(2008{\natexlab{a}})}]{2008GeoRL..35.9201R}
{Roberts}, J.~H., {Nimmo}, F., may 2008{\natexlab{a}}. {Near-surface heating on
  Enceladus and the south polar thermal anomaly}. Geophysical Research Letters
  35, 9201.

\bibitem[{{Roberts} and {Nimmo}(2008{\natexlab{b}})}]{roberts2008near}
{Roberts}, J.~H., {Nimmo}, F., 2008{\natexlab{b}}. {Near-surface heating on
  Enceladus and the south polar thermal anomaly}. Geophysical Research Letters
  35~(9).

\bibitem[{{Rothschild} et~al.(1994){Rothschild}, {Giver}, {White}, and
  {Mancinelli}}]{rothschild1994metabolic}
{Rothschild}, L.~J., {Giver}, L.~J., {White}, M.~R., {Mancinelli}, R.~L., 1994.
  {METABOLIC ACTIVITY OF MICROORGANISMS IN EVAPORITES1}. Journal of Phycology
  30~(3), 431--438.

\bibitem[{{Rothschild} and {Mancinelli}(2001)}]{rothschild2001life}
{Rothschild}, L.~J., {Mancinelli}, R.~L., 2001. {Life in extreme environments}.
  Nature 409~(6823), 1092--1101.

\bibitem[{{Sassen} et~al.(2004){Sassen}, {Roberts}, {Carney}, {Milkov},
  et~al.}]{sassen2004free}
{Sassen}, R., {Roberts}, H.~H., {Carney}, R., {Milkov}, A.~V., et~al., 2004.
  {Free hydrocarbon gas, gas hydrate, and authigenic minerals in chemosynthetic
  communities of the northern Gulf of Mexico continental slope: relation to
  microbial processes}. Chemical Geology 205~(3), 195--217.

\bibitem[{{Saur} et~al.(2008){Saur}, {Schilling}, {Neubauer}, {Strobel},
  {Simon}, {Dougherty}, {Russell}, and {Pappalardo}}]{2008GeoRL..3520105S}
{Saur}, J., {Schilling}, N., {Neubauer}, F.~M., {Strobel}, D.~F., {Simon}, S.,
  {Dougherty}, M.~K., {Russell}, C.~T., {Pappalardo}, R.~T., oct 2008.
  {Evidence for temporal variability of Enceladus' gas jets: Modeling of
  Cassini observations}. Geophysical Research Letters 35, L20105.

\bibitem[{{Schmidt} et~al.(2008){Schmidt}, {Brilliantov}, {Spahn}, and
  {Kempf}}]{2008Natur.451..685S}
{Schmidt}, J., {Brilliantov}, N., {Spahn}, F., {Kempf}, S., feb 2008. {Slow
  dust in Enceladus' plume from condensation and wall collisions in tiger
  stripe fractures}. Nature 451, 685--688.

\bibitem[{Schneider et~al.(2009)Schneider, Burger, Schaller, Brown, Johnson,
  Kargel, Dougherty, and Achilleos}]{schneider2009no}
Schneider, N.~M., Burger, M.~H., Schaller, E.~L., Brown, M.~E., Johnson, R.~E.,
  Kargel, J.~S., Dougherty, M.~K., Achilleos, N.~A., 2009. {No sodium in the
  vapour plumes of Enceladus}. Nature 459~(7250), 1102--1104.

\bibitem[{{Schubert} et~al.(2007){Schubert}, {Anderson}, {Travis}, and
  {Palguta}}]{2007Icar..188..345S}
{Schubert}, G., {Anderson}, J.~D., {Travis}, B.~J., {Palguta}, J., 2007.
  {Enceladus: Present internal structure and differentiation by early and
  long-term radiogenic heating}. Icarus 188~(2), 345--355.

\bibitem[{{Seckbach}(2013)}]{seckbach2013enigmatic}
{Seckbach}, J., 2013. {Enigmatic microorganisms and life in extreme
  environments}. Vol.~1. Springer Science \& Business Media.

\bibitem[{{Sekine} et~al.(2015){Sekine}, {Shibuya}, {Postberg},
  et~al.}]{2015NatCo...6E8604S}
{Sekine}, Y., {Shibuya}, T., {Postberg}, F., et~al., oct 2015.
  {High-temperature water-rock interactions and hydrothermal environments in
  the chondrite-like core of Enceladus}. Nature Communications 6, 8604.

\bibitem[{{Simon} et~al.(2011){Simon}, {Saur}, {Kriegel}, {Neubauer},
  {Motschmann}, and {Dougherty}}]{2011JGRA..116.4221S}
{Simon}, S., {Saur}, J., {Kriegel}, H., {Neubauer}, F.~M., {Motschmann}, U.,
  {Dougherty}, M.~K., apr 2011. {Influence of negatively charged plume grains
  and hemisphere coupling currents on the structure of Enceladus' Alfv{\'e}n
  wings: Analytical modeling of Cassini magnetometer observations}. Journal of
  Geophysical Research (Space Physics) 116, A04221.

\bibitem[{{Smith} et~al.(1982){Smith}, {Soderblom}, {Batson},
  et~al.}]{1982Sci...215..504S}
{Smith}, B.~A., {Soderblom}, L., {Batson}, R.~M., et~al., jan 1982. {A new look
  at the Saturn system - The Voyager 2 images}. Science 215, 504--537.

\bibitem[{{Sotin} et~al.(2011){Sotin}, {Altwegg}, {Brown}, {Hand}, {Lunine},
  {Soderblom}, {Spencer}, {Tortora}, and {JET Team}}]{2011LPI....42.1326S}
{Sotin}, C., {Altwegg}, K., {Brown}, R.~H., {Hand}, K., {Lunine}, J.~I.,
  {Soderblom}, J., {Spencer}, J., {Tortora}, P., {JET Team}, mar 2011. {JET:
  Journey to Enceladus and Titan}. In: Lunar and Planetary Science Conference.
  Vol.~42 of Lunar and Planetary Inst. Technical Report. p. 1326.

\bibitem[{{Spahn} et~al.(2006){Spahn}, {Schmidt}, {Albers},
  et~al.}]{2006Sci...311.1416S}
{Spahn}, F., {Schmidt}, J., {Albers}, N., et~al., mar 2006. {Cassini Dust
  Measurements at Enceladus and Implications for the Origin of the E Ring}.
  Science 311, 1416--1418.

\bibitem[{{Spencer} and {Nimmo}(2013)}]{2013AREPS..41..693S}
{Spencer}, J.~R., {Nimmo}, F., may 2013. {Enceladus: An Active Ice World in the
  Saturn System}. Annual Review of Earth and Planetary Sciences 41, 693--717.

\bibitem[{{Spencer} et~al.(2006){Spencer}, {Pearl}, {Segura},
  et~al.}]{2006Sci...311.1401S}
{Spencer}, J.~R., {Pearl}, J.~C., {Segura}, M., et~al., mar 2006. {Cassini
  Encounters Enceladus: Background and the Discovery of a South Polar Hot
  Spot}. Science 311, 1401--1405.

\bibitem[{{Spitale} et~al.(2015){Spitale}, {Hurford}, {Rhoden}, {Berkson}, and
  {Platts}}]{2015Natur.521...57S}
{Spitale}, J.~N., {Hurford}, T.~A., {Rhoden}, A.~R., {Berkson}, E.~E.,
  {Platts}, S.~S., may 2015. {Curtain eruptions from Enceladus' south-polar
  terrain}. Nature 521, 57--60.

\bibitem[{{Spitale} and {Porco}(2007)}]{2007Natur.449..695S}
{Spitale}, J.~N., {Porco}, C.~C., oct 2007. {Association of the jets of
  Enceladus with the warmest regions on its south-polar fractures}. Nature 449,
  695--697.

\bibitem[{{Stevenson} et~al.(2015){Stevenson}, {Burkhardt}, {Cockell},
  et~al.}]{stevenson2015multiplication}
{Stevenson}, A., {Burkhardt}, J., {Cockell}, C.~S., et~al., 2015.
  {Multiplication of microbes below 0.690 water activity: implications for
  terrestrial and extraterrestrial life}. Environmental microbiology 17~(2),
  257--277.

\bibitem[{{Strange} et~al.(2009){Strange}, {Campagnola}, and
  {Russell}}]{strange2009leveraging}
{Strange}, N.~J., {Campagnola}, S., {Russell}, R.~P., 2009. Leveraging flybys
  of low mass moons to enable an enceladus orbiter. Advances in the
  Astronautical Sciences 135~(3), 2207--2225.

\bibitem[{{Thomas} et~al.(2016){Thomas}, {Tajeddine}, {Tiscareno}, {Burns},
  {Joseph}, {Loredo}, {Helfenstein}, and {Porco}}]{2016Icar..264...37T}
{Thomas}, P.~C., {Tajeddine}, R., {Tiscareno}, M.~S., {Burns}, J.~A., {Joseph},
  J., {Loredo}, T.~J., {Helfenstein}, P., {Porco}, C., jan 2016. {Enceladus's
  measured physical libration requires a global subsurface ocean}. Icarus 264,
  37--47.

\bibitem[{{Tian} et~al.(2007){Tian}, {Stewart}, {Toon}, {Larsen}, and
  {Esposito}}]{2007Icar..188..154T}
{Tian}, F., {Stewart}, A.~I.~F., {Toon}, O.~B., {Larsen}, K.~W., {Esposito},
  L.~W., may 2007. {Monte Carlo simulations of the water vapor plumes on
  Enceladus}. Icarus 188, 154--161.

\bibitem[{{Tobie} et~al.(2014){Tobie}, {Teanby}, {Coustenis}, {Jaumann},
  {Raulin}, {Schmidt}, {Carrasco}, {Coates}, {Cordier}, {De Kok}, {Geppert},
  {Lebreton}, {Lefevre}, {Livengood}, {Mandt}, {Mitri}, {Nimmo}, {Nixon},
  {Norman}, {Pappalardo}, {Postberg}, {Rodriguez}, {Schulze-Makuch},
  {Soderblom}, {Solomonidou}, {Stephan}, {Stofan}, {Turtle}, {Wagner}, {West},
  and {Westlake}}]{2014P&SS..104...59T}
{Tobie}, G., {Teanby}, N.~A., {Coustenis}, A., {Jaumann}, R., {Raulin}, F.,
  {Schmidt}, J., {Carrasco}, N., {Coates}, A.~J., {Cordier}, D., {De Kok}, R.,
  {Geppert}, W.~D., {Lebreton}, J.-P., {Lefevre}, A., {Livengood}, T.~A.,
  {Mandt}, K.~E., {Mitri}, G., {Nimmo}, F., {Nixon}, C.~A., {Norman}, L.,
  {Pappalardo}, R.~T., {Postberg}, F., {Rodriguez}, S., {Schulze-Makuch}, D.,
  {Soderblom}, J.~M., {Solomonidou}, A., {Stephan}, K., {Stofan}, E.~R.,
  {Turtle}, E.~P., {Wagner}, R.~J., {West}, R.~A., {Westlake}, J.~H., dec 2014.
  {Science goals and mission concept for the future exploration of Titan and
  Enceladus}. Planetary and Space Science 104, 59--77.

\bibitem[{{Tobie} et~al.(2008){Tobie}, {{\v C}adek}, and
  {Sotin}}]{2008Icar..196..642T}
{Tobie}, G., {{\v C}adek}, O., {Sotin}, C., aug 2008. {Solid tidal friction
  above a liquid water reservoir as the origin of the south pole hotspot on
  Enceladus}. Icarus 196, 642--652.

\bibitem[{{Tsou} et~al.(2012){Tsou}, {Brownlee}, {McKay}, {Anbar}, {Yano},
  {Altwegg}, {Beegle}, {Dissly}, {Strange}, and {Kanik}}]{2012AsBio..12..730T}
{Tsou}, P., {Brownlee}, D.~E., {McKay}, C.~P., {Anbar}, A.~D., {Yano}, H.,
  {Altwegg}, K., {Beegle}, L.~W., {Dissly}, R., {Strange}, N.~J., {Kanik}, I.,
  aug 2012. {LIFE: Life Investigation For Enceladus A Sample Return Mission
  Concept in Search for Evidence of Life}. Astrobiology 12, 730--742.

\bibitem[{{Waite} et~al.(2006){Waite}, {Combi}, {Ip},
  et~al.}]{2006Sci...311.1419W}
{Waite}, J.~H., {Combi}, M.~R., {Ip}, W.~H., et~al., mar 2006. {Cassini Ion and
  Neutral Mass Spectrometer: Enceladus Plume Composition and Structure}.
  Science 311, 1419--1422.

\bibitem[{{Waite} et~al.(2004){Waite}, {Lewis}, {Kasprzak},
  et~al.}]{waite2004cassini}
{Waite}, J.~H., {Lewis}, W.~S., {Kasprzak}, W.~T., et~al., 2004. {The Cassini
  ion and neutral mass spectrometer (INMS) investigation}. In: The
  Cassini-Huygens Mission. Springer, pp. 113--231.

\bibitem[{{Waite} et~al.(2009){Waite}, {Lewis}, {Magee},
  et~al.}]{2009Natur.460..487W}
{Waite}, J.~H., {Lewis}, W.~S., {Magee}, B.~A., et~al., jul 2009. {Liquid water
  on Enceladus from observations of ammonia and $^{40}$Ar in the plume}. Nature
  460, 487--490.

\bibitem[{{Zolotov}(2007)}]{2007GeoRL..3423203Z}
{Zolotov}, M.~Y., dec 2007. {An oceanic composition on early and today's
  Enceladus}. Geophysical Research Letters 34, L23203.

\end{thebibliography}

\end{document}